\documentclass[12pt]{article}
\usepackage{amsmath,amsfonts,amssymb}
\usepackage{mathtools}
\usepackage{graphicx}
\usepackage{algorithm}
\usepackage{comment}
\usepackage{setspace}
\usepackage{algpseudocode}
\RequirePackage[authoryear]{natbib}
\RequirePackage[colorlinks,citecolor=blue,urlcolor=blue]{hyperref}
\usepackage{multirow}
\usepackage{caption, array, float}
\usepackage{subcaption, booktabs}
\usepackage{bm}
\usepackage[left=1in,right=1in,top=1.5in,bottom=1.5in]{geometry}
\usepackage{bbm}
\usepackage{xcolor}
\usepackage{placeins}
\usepackage{authblk}
\usepackage{setspace}
\usepackage{tikz}
\newcommand{\revision}[1]{{\leavevmode\color{black}#1}}

\DeclareMathOperator*{\arginf}{arg\,inf}
\usepackage{amsthm}
\newtheorem{theorem}{Theorem}[section]

\newtheorem{lemma}[theorem]{Lemma}
\date{}
\newcolumntype{H}{>{\setbox0=\hbox\bgroup}c<{\egroup}@{}}
\usepackage{geometry}
\title{Enhancing Empathic Accuracy: Penalized Functional Alignment Method to Correct Temporal Misalignment in Real-time Emotional Perception}

\begin{document}

\author[1]{Linh H. Nghiem}
\author[2]{Chrystyna Kouros}
\author[3]{Jing Cao} 
\author[3]{Chul Moon} 
\affil[1]{School of Mathematics and Statistics, University of Sydney, Australia}
\affil[2]{Department of Psychology, Southern Methodist University, USA}
\affil[3]{Department of Statistics and Data Science, Southern Methodist University, USA}

\maketitle

\begin{abstract}
Empathic accuracy (EA) is the ability to accurately understand thoughts and feelings of another person, which is crucial for social and psychological interactions. Traditionally, EA is assessed by comparing a perceiver\textquotesingle s moment-to-moment ratings of a target\textquotesingle s emotional state with the target\textquotesingle s own self-reported ratings at corresponding time points. However, misalignments between these two sequences are common due to the complexity of emotional interpretation and individual differences in behavioral responses. Conventional methods often ignore or oversimplify these misalignments, for instance by assuming a fixed time lag, which can introduce bias into EA estimates.  To address this, we propose a novel alignment approach that captures a wide range of misalignment patterns. Our method leverages the square-root velocity framework to decompose emotional rating trajectories into amplitude and phase components. To ensure realistic alignment, we introduce a regularization constraint that limits temporal shifts to ranges consistent with human perceptual capabilities. This alignment is efficiently implemented using a constrained dynamic programming algorithm. We validate our method through simulations and real-world applications involving video and music datasets, demonstrating its superior performance over traditional techniques.
\end{abstract}
\section{Introduction}
	\label{sec:int}

 The ability to perceive and understand the emotions and thoughts of others, broadly referred to as empathy, plays an important role in human society by facilitating cooperation and social cohesion \citep{de2008putting}.  While empathy encompasses multiple components, including sharing in another's emotional experience and concern for others, empathic accuracy (EA) refers to the specific skill of accurately inferring what another person is thinking and feeling in a given moment \citep{Ickes1997}. EA is typically measured behaviorally by comparing a perceiver's rating of a target's emotional state to the target's own self-reported emotional experience. 
 % On a personal level, empathy is essential for social interactions. Impairments in empathy, such as those observed in autism and psychopathy, lead to significant social dysfunction \citep{Blair2005}.To investigate the variability of empathy among individuals, researchers use empathic accuracy (EA) to measure how accurately people can infer the specific thoughts and feelings of others. EA essentially quantifies the degree of empathy by comparing a perceiver's judgment of a target's emotional state to the target's self-reported emotional experience. 
 Given its importance to social interactions and quality of life, EA has become a focal point of research across various fields.  For example, in social science, EA’s role was examined in developing and maintaining healthy social relationships \citep{sened2017empathic}. In clinical research, EA has been used as an index to differentiate individuals with certain psychiatric disorders from healthy controls \citep{lee2011schizophrenia}. However, the validity of these studies critically depends on the quality and accuracy of EA measurement.

%Two types of studies are commonly used to examine EA. One is the non-real time EA study, where perceivers provide their response to stimuli after the stimuli have been conducted. The outcome on their empathy can be categories of emotion (e.g., happiness, anger, or sadness, etc.) or extent of emotion on a Likert-type scale \citep{ekman1992argument,schweinle2002emphatic}. However, empathy fluctuates simultaneously due to various factors such as internal state changes or in response to varying external factors \citep{ashar2017empathic,mackes2018tracking}. On the other hand, correlational analysis and its monotonic transformation are the conventional and arguably the most common statistical method analyzing EA data \citep{zaki2009unpacking,mackes2018tracking}. In a broader context of modeling the accuracy of human judgment, a number of approaches have been proposed as an alternative for correlation analysis, such as the Truth and Bias Model \citep{west2011truth} and the Social Accuracy Model \citep{biesanz2010social}. 
%Both take the form of a linear model, where the former assumes that judgments on the subject are pulled by 2 primary forces, the truth force (i.e., the truth or reality) and the bias force (i.e.,  the extent to which judgments are being pulled toward bias values), and the latter estimates perceiver and target effects in interpersonal perception such as how we see others and are seen by others.

\begin{figure}[!ht]
    \centering
    \includegraphics[width=0.7\linewidth]{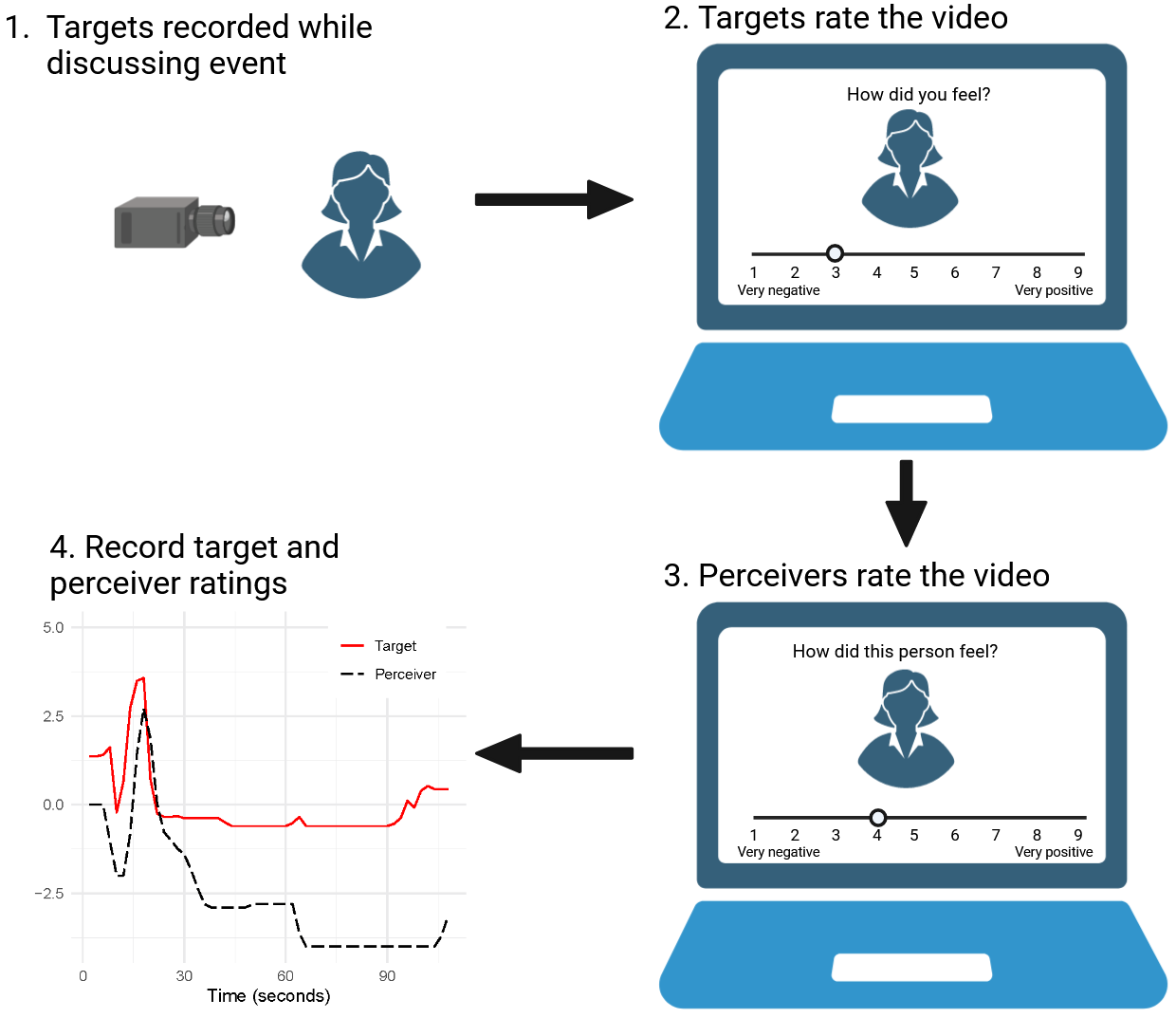}
    \caption{Example of real-time EA data collection procedure.}
    \label{figure: real-time EA}
\end{figure}

There are two types of studies commonly used to examine EA. One is the non-real-time EA study design, where perceivers provide their response to stimuli after the stimuli have been conducted. The outcome of their overall empathy can be the categories of emotion (e.g., happiness, anger, sadness, etc.) or the extent of emotion on a Likert-type scale \citep{ekman1992argument,schweinle2002emphatic}. The other EA study design is the real-time assessment of perceivers' empathy on an audio or video stimuli \revision{without pausing} (i.e., the recorded affective states of targets) \citep{zaki2008takes, JOSPE2020296}, where perceivers provide continuous feedback on their perceptions of the target's emotional state while the stimuli is unfolding. Illustrated in Figure~\ref{figure: real-time EA}, social targets varying in trait emotional intensity were videotaped while discussing emotional autobiographical events. Perceivers watch these videos and report the perceived emotions every two seconds using, for example, a 9-point Likert scale (e.g., 1 = extremely negative; 9 = extremely positive). Compared with the non-real-time EA studies, the real-time design provides more granular information on the dynamic nature of empathy in everyday interactions and detects subtle changes in emotional responses that might be missed in non-real-time assessments. 

In this paper, we focus on analyzing data from real-time EA study designs. For such designs, correlational analysis \citep[among others]{mackes2018tracking,zaki2009unpacking} is a predominant statistical method for examining EA. This approach computes a monotonic transformation of the Pearson correlation between the observed perceivers’ responses with targets’ self-reported emotion rating.
%, where the latter is considered the gold standard. 
Linear models have also been introduced to investigate the influence of additional factors or unobserved variables on EA. For example, \citet{tabak2022initial} proposed a latent variable model that decomposes EA into three separate dimensions: bias, discrimination, and variability. Bias measures the systematic difference between perceiver's ratings and target's ratings; discrimination measures perceiver’ sensitivity in relation to target's ratings; and variability measures the variance of random error in perceiver’s perceptions. 

A key assumption in traditional correlational and linear model analyses of empathic accuracy is that perceivers’ and targets’ rating sequences are perfectly aligned—that is, a perceiver’s rating at a given time point is directly compared to the target’s rating at that same moment.  However, this assumption often fails in practice due to the complex cognitive processes involved in interpreting another person’s emotional state and the time required to produce a behavioral response.  Scherer's multi-stage model of emotion decoding \cite{scherer2003vocal} highlights how perceivers actively interpret dynamic cues such as facial expressions, gestures, and vocalizations to infer emotions, a process that naturally introduces temporal delays. Additionally, the act of recording a response, such as pressing a key or moving a joystick, can vary in duration, further contributing to misalignment. To address these issues, we posit that each perceiver has an underlying latent rating that reflects their true empathic understanding, independent of these timing distortions.

\begin{figure}[!ht]
    \centering
    \begin{tikzpicture}[every node/.style={align=center}]
  % Nodes
  \node (target) at (0,3) {target's\\rating};
  \node (latent) at (2,1.5) {perceiver's\\latent rating};
  \node (observed) at (0,0) {perceiver's\\observed rating};
  % Arrows
  \draw[<->] (observed) -- node[midway, left] {\textbf{A}} (target);
  \draw[<->] (latent) -- node[midway, above right] {\textbf{B}} (target);
  \draw[<->] (observed) -- node[midway, below right] {\textbf{C}} (latent);
\end{tikzpicture}
    \caption{Illustration of different relations between a target’s rating, a perceiver’s latent rating, and the perceiver’s observed rating.  Discrepancy A denotes the difference between a target’s rating and a perceiver’s observed rating. Disagreement B captures the inconsistency between target’s rating and perceiver’s latent rating, which is the true focus of 
    EA. Misalignment C refers to the divergence between perceiver’s observed rating and their latent rating, often due to distortions in expressing their internal judgment. Discrepancy A can arise from both Disagreement B and Misalignment C. Most conventional EA methods mistakenly assess Discrepancy A, thereby conflating measurement error with genuine empathic inaccuracy.}
    \label{fig:enter-label}
\end{figure}

Figure~\ref{fig:enter-label} illustrates the relations between a target’s rating, a perceiver’s latent rating, and the perceiver’s observed rating. 
Discrepancy A, the difference between the target’s rating and the perceiver’s observed rating, can arise from two sources: Disagreement B, which reflects the true empathic inaccuracy between the target’s rating and the perceiver’s latent rating, and Misalignment C, which captures the temporal mismatch between the perceiver’s latent and observed ratings. Crucially, EA is intended to measure a perceiver’s ability to correctly infer another person’s emotional state—that is, to quantify Disagreement B—not the speed or timing of their response. A perceiver who accurately identifies a target’s emotion, even with a slight misalignment, should not be penalized as less empathically accurate. However, traditional EA methods often ignore Misalignment C, relying solely on comparisons between the target’s and perceiver’s observed ratings (i.e., measuring Discrepancy A). This could potentially lead to biased estimates of EA and inflated variability due to even minor timing discrepancies. Our work addresses this limitation by explicitly correcting for Misalignment C, thereby yielding more accurate estimates of Disagreement B and preserving the psychological validity of EA assessments.

\begin{figure}[!ht]
    \centering
    \begin{subfigure}{0.47\linewidth}
    \centering
    \includegraphics[width=\linewidth]{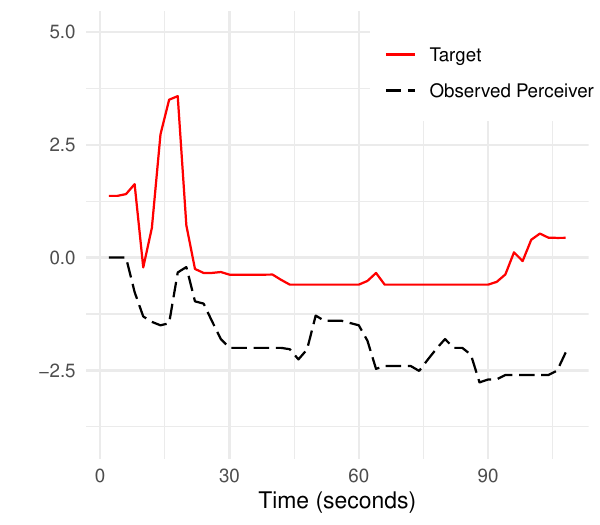}
    \caption{Emotion rating functions}
    \label{subfig:DelvinFns}
    \end{subfigure}
    \begin{subfigure}{0.47\linewidth}
    \centering
    \includegraphics[width=\linewidth]{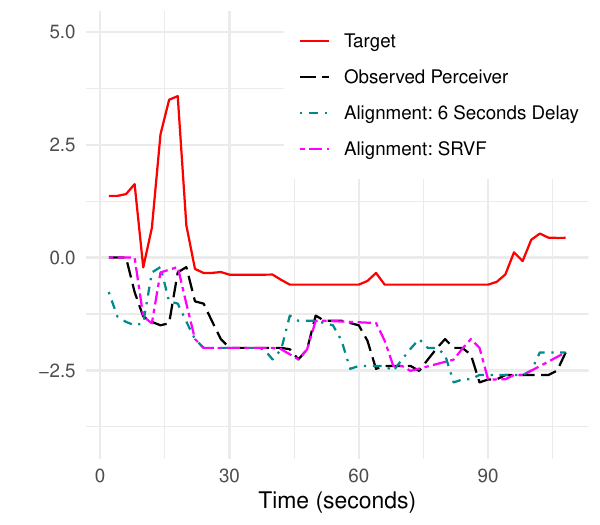}
    \caption{Aligned functions}
    \label{subfig:DelvinAligned}
    \end{subfigure}
    \caption{(a) An example of misaligned rating sequences between a perceiver and a target. The solid red line represents the target's ratings
    and the black dashed perceiver's ratings. (b)  Aligned ratings for the perceiver. The green dashed line shows the 6-second delay adjustment, and the purple dashed line shows the aligned ratings using the penalized SRVF representation.}
\label{fig:delvin}
\end{figure}

Note that common approaches to address misalignment in EA studies involve introducing a fixed response delay, assuming consistent emotional expression patterns across individuals. This method shifts perceivers' response time series backward by a predetermined amount \citep{nicolle2012robust,huang2015investigation,khorram2019jointly}. However, \cite{scherer2003vocal} countered this assumption, arguing that emotional expressions are diverse and context-dependent.  In addition, a review of event-related potential (ERP) studies spanning 40 years found that emotional stimuli elicit differences in neural processing speed based on valence and arousal level \cite{Olofsson2008}. The findings suggest that stimuli with higher motivational relevance receive priority in neural processing. Consequently, the misalignment between perceivers' and targets' ratings is more complex than a simple, fixed time shift applied to all participants. 
%given that stimuli in EA studies are dynamic where emotional valence and intensity fluctuate continuously, 
It would be inappropriate to apply a fixed time delay in EA studies, as it fails to account for the variability in emotional processing across different moments. Figure~\ref{subfig:DelvinFns}
illustrates an example of misalignment between perceiver and target ratings in an EA study \citep{devlin2014not}. While a delay in perceiver responses is evident, it is not the sole cause of misalignment. For instance, the perceiver's prolonged sustained response from 10 to 15 seconds, in contrast to the target's brief dip at 10 seconds, highlights the complex nature of these discrepancies.

To accommodate a wider range of misalignment patterns beyond simple time shifts, time series alignment methods aim to preserve key structural features in the data, such as peaks and valleys, ensuring more accurate analysis. One widely used technique is dynamic time warping (DTW), which aligns time series by stretching or compressing the time axis to match similar patterns \citep{sakoe1978dynamic,berndt1994using}. However, DTW can sometimes introduce distortions by forcing unnatural alignments between sequences \citep{srivastava2011registration,marron2015functional,zhao2020modeling}. To mitigate this, smoothness penalties have been proposed \citep{ramsay2005functional}. However, it may also lead to biased alignments \citep{guo2022data}. Alternatively, landmark-based methods align time series by identifying and matching distinctive features like peaks and valleys \citep{kneip2000curve}. While potentially effective, these methods are highly sensitive to noise and may lose important information due to the discretization of continuous functions into a limited set of landmarks \citep{wang1997alignment,marron2015functional}. Moreover, such approaches are ill-suited for real-time emotion rating data in EA studies, where there is no clear consensus on the number or the location of meaningful landmarks.

%As for the landmark-based alignment method, detecting common landmarks from noisy data is challenging \citep{wang1997alignment}, and discretization of functions using a limited number of landmarks could lead to a loss of information \citep{marron2015functional}.
%In addition, deep learning methods that implement temporal transformer layers have been studied \citep{oh2018learning,lohit2019temporal}.

Due to the high-frequency nature of the observed EA rating data, we treat each observed curve as a sample path of a continuous function in the time domain, i.e., functional data. Such an approach of representing high-frequency data as functional is common in the literature %\citep[Chapter 2]{kokoszka2017introduction}. 
\citep{kokoszka2017introduction}. 
From this perspective, misalignment between two observed ratings could be explained by a smooth warping function that distorts the time domain of the perceiver relative to that of the target. Hence, the target and the perceiver's rating functions can be aligned by estimating this smooth warping function from the observed data, for example, by minimizing an $\mathbb{L}^2$ distance between the target and the estimated aligned response function \citep{ramsay1998curve}. Recently, the square root velocity function (SRVF) representation has been employed for aligning functions \citep{srivastava2011registration}, and has been increasingly applied across various fields, including biology, medicine, geology, and signal processing \citep{su2014statistical, laga2014landmark, bharath2018radiologic, zhao2020modeling, mitchell2022object}. As we will review in Section \ref{sec:background}, this SRVF representation leverages the Fisher-Rao metric's invariance property, and enables a consistent separation of horizontal component (also known as phase) from vertical component (also known as amplitude) of functions, making visualization and summarizing variability in functional datasets more effective \citep{xie2017geometric}.

Building upon the SRVF-alignment framework, this paper introduces a novel penalized SRVF-based alignment method for unsynchronized rating sequences in EA studies. Our approach introduces both practical and methodological innovations. Practically, it is the first method in EA research to accommodate a wide range of misalignment patterns (e.g., delays, compressions, and stretches), moving beyond the limitations of fixed-delay adjustments. Methodologically, we incorporate a novel penalty term that constrains temporal shifts within bounds consistent with human perceptual capabilities, thereby preventing excessive or unrealistic alignments \citep{levenson1988emotion,gunes2010automatic,ringeval2015prediction,mariooryad2014correcting}. This is important because not all temporal discrepancies should be corrected; some may reflect genuine empathic inaccuracy rather than misalignment. To address this, our penalized alignment method selectively adjusts only short-term misalignments—those occurring within a few seconds—treating them as Misalignment C (as shown in Figure~\ref{fig:enter-label}). In contrast, larger discrepancies, which may indicate a lack of empathic understanding (Disagreement B), are preserved.

To highlight the contribution of our method, Figure~\ref{subfig:DelvinAligned} compares the proposed penalized SRVF method with a fixed 6-second delay adjustment. Although the 6-second delay adjustment aligns the peaks between the two sequences, it, unfortunately, eliminates the brief 5-second sustain at the start of the perceiver's sequence, which originally matched up with the target's self-rating sequence. In contrast, the proposed penalized SRVF-based method has aligned the peaks while keeping the initial sustain in the perceiver's sequence in place, demonstrating its flexibility in handling complex misalignment patterns. By enabling a more precise alignment, our method yields a more accurate estimation of EA, avoiding the pitfalls of underestimation when misalignment is ignored and overestimation when no penalty is applied.

The remainder of the paper is structured as follows. Section 2 provides background information on empathic accuracy and existing alignment methods. The proposed methodology is detailed in Section 3. To evaluate the proposed method, Section 4 presents a simulation study and comparisons to alternative approaches. Real-world applications of assessing EA in social and music contexts are explored in Section 5. Finally, Section 6 offers a discussion of the findings and concludes the paper.

\section{Background}
\label{sec:background}
\subsection{Elastic Functional Data Analysis}

Functional data often exhibit both vertical and horizontal differences, where the latter is known as phase variation and characterized by misaligned geometric features such as peaks and valleys in the time domain \citep{wu2014analysis,wallace2014pairwise,tucker2014analysis}
Let $x,a,y:[t_0,t_T]\rightarrow \mathbb{R}$ be the function of the target rating, the function of the perceiver's latent rating, and the function of the perceiver's observed rating, respectively. To account for both vertical and horizontal difference between these two functions, we assume a data generation process that $y(t) = f(x(\tilde{t}), \varepsilon(\tilde{t}))$ where $\tilde{t}=\psi(t)$,  $f:\mathbb{R}^2 \to \mathbb{R}$ is a link function, $\psi:[t_0, t_T] \to [t_0, t_T]$ is a time warping function, $\circ$ denotes the composition operator, and $\varepsilon$ is a random noise function. Essentially, the process of generating the perceiver's observed rating $y$ is decomposed into a transformation step and a warping step, as depicted in the following expression \eqref{fig:DGF}. 
\begin{equation}
\begin{tikzpicture}[baseline=(current bounding box.center),->,>=stealth, thick]
  \node (A) at (0,0) {$x$};
  \node (B) at (3.5,0) {$a = f(x, \epsilon)$};
  \node (C) at (7.5,0) {$y = a \circ \psi$};

  \draw (A) -- (B) node[midway, above]{Transform};
  \draw (B) -- (C) node[midway, above] {Warp};

%  \node at (1.5,-0.3) {\text{vertical + diff}};
%\node at (5.5,-0.3) {\text{horizontal diff}};
\end{tikzpicture}
\label{fig:DGF}
\end{equation}

Note that in this transformation step, the link  $f$ matches the target function $x$ at a time point $t$ to the perceiver's latent function $a$ at the same time $t$. This correspondence is distorted by a warping function $\psi$ in the second warping step, so the target function $x$ at time $t$ now is matched to the perceiver's observed function $y$ at $\psi(t)$. The warping function $\psi$ is usually assumed to belong to the set of smoothing functions $\Gamma_I$, where
$$
\Gamma_I = \{\psi \mid \psi(t_0)=t_0, \psi(t_T)=t_T, ~ \psi^\prime(t) ~\text{exists}~, \psi^\prime(t) \geq 0, ~ \gamma = \psi^{-1} \text{exists} \}.
$$

Relating this process to Figure \ref{fig:enter-label}, the transformation step models the Disagreement B, while the warping step models the Misalignment C, and the Discrepancy A accumulates both steps together. In the context of measuring EA, the perceiver's latent function $a$ represents a rating from the perceiver that is aligned with $x$, i.e., that \emph{can be compared with the target $x$ point-to-point in time},  and a measure of EA is a similarity measure between $x$ and $a$.  

The data generation process \eqref{fig:DGF} motivates the following workflow for quantifying EA. Because $y = a \circ \psi$, we can write $a = y \circ\gamma$, where \revision{the inverse warping function} $\gamma = \psi^{-1}\revision{\in \Gamma_I}$ is assumed to exist since $\psi \in \Gamma_I$. Hence,
 we first conduct an alignment step to obtain an estimated inverse warping function $\hat{\gamma}$ and an estimated latent function $\hat{y} = {y} \circ \hat\gamma$ from the observed target and perceiver functions $x$ and  $y$. Then, we could estimate EA by a similarity measure between $x$ and $\hat{y}$. 
 
 Since misalignment between two functions is inherently related to the difference in how fast they move, a common way to conduct the alignment step is to compare how these functions change over time, which is mathematically described by their corresponding first derivative. Therefore, the general idea of the SRVF-based alignment methods is to minimize the distance between the first derivative of the target $x$ and the estimated function $\hat{y}$.

We briefly review the formulation of the SRVF representation here, where more details can be found in \citet{srivastava2016functional}. For any absolute continuous function $f:[t_0,t_T]\rightarrow \mathbb{R}$, the SRVF of $f$ is the function $q_f:[t_0,t_T] \rightarrow \mathbb{R}$,
$q_f(t) = \text{sign}\left\{{f}^\prime(t)\right\}\sqrt{|f^\prime(t)|}$, where ${f}^\prime(t)=df/dt$. As described in the previous paragraph $q_f$, this SRVF is defined based on the first derivative $f^\prime$; the specific form of $q_f(t)$ is motivated to keep its norm unaffected by the warping, which is useful to separate a function into its amplitude and phase component \citep{srivastava2016functional}. Specifically, if $f$ is warped by $\gamma$, the corresponding SRVF of $f\circ\gamma$ becomes $q_{f\circ\gamma} = (q_f\circ\gamma)\sqrt{\gamma^\prime}$, but \revision{the squared $\mathbb{L}^2$ norm is preserved} $\| (q_{f\circ\gamma}) \|_2^2 = \|q_f\|_2^2$.    Let $q_x$ and $q_y$ be the SRVFs of the target and perceiver functions, respectively. 
Then, the SRVF-based alignment method aims to find an optimal \revision{inverse} warping function that minimizes this discrepancy, i.e.,
\begin{equation}
    \hat{\gamma}_{u} = \arginf_{\gamma \in  \revision{\Gamma_I} } \| q_x-(q_y, \gamma) \|_2^2,
    \label{eq:unpenwarpsrvf}
\end{equation} 
where we write $q_{f\circ\gamma} = (q_f, \gamma)$ to ease the notation. The optimal $\hat{\gamma}_{u}$ is expected to align two functions so that the transformed function $\hat{y} = y \circ \hat{\gamma}_{u}$ is aligned with $x$. 
The subscript ${u}$ stands for ``unpenalized,'' meaning the optimal $\hat{\gamma}_{u}$ is not subject to any other constraint than being in the space $\Gamma_I$. This unpenalized alignment has been implemented in the  \texttt{fdasrvf} packages \citep{tucker2023fdasrvf} in both R and Python. After conducting the alignment step, in addition to the EA measure obtained  by computing a similarity metric between $\hat{y}$ and $x$, we can also quantify the amount of warping made by each perceiver in relative to the target by a Fisher-Rao phase distance between the estimated warping $\hat\gamma_u$ and the identity warping $\gamma_{id}(t) = t $ as

\begin{equation}
    d_p(x,y) \approx \cos^{-1}\left(\int_0^1 \sqrt{{\hat{\gamma}^\prime}_{u}(t)}~dt\right),
    \label{eq:fisherrao}
\end{equation}
\noindent which is a proper metric distance on the set \revision{$\Gamma_I$} \citep{srivastava2016functional}.

\begin{figure}[!ht]
    \centering
    \begin{subfigure}{1\linewidth}
    \centering
    \includegraphics[width=0.42\linewidth]{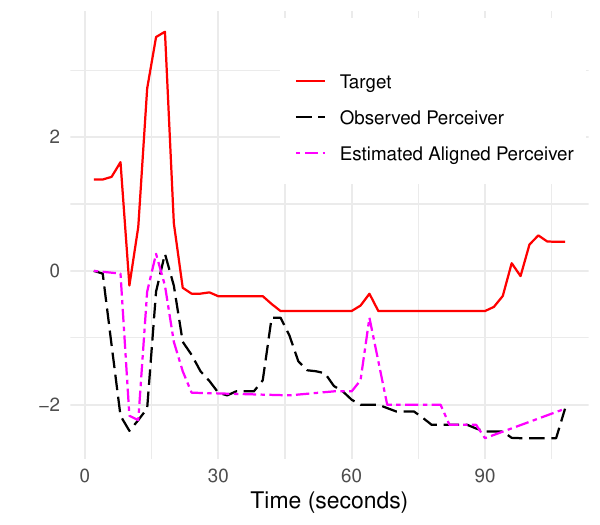}
    \includegraphics[width=0.42\linewidth]{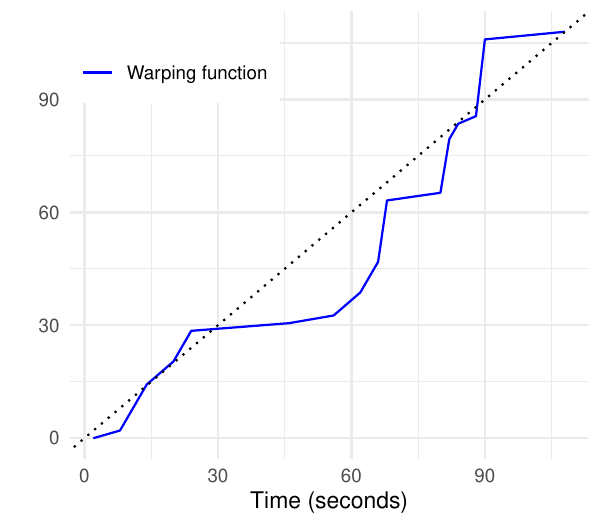}
    \caption{Unpenalized alignment}
    \label{subfig:DelvinFnunpen}
    \end{subfigure}
    \begin{subfigure}{1\linewidth}
    \centering
    \includegraphics[width=0.42\linewidth]{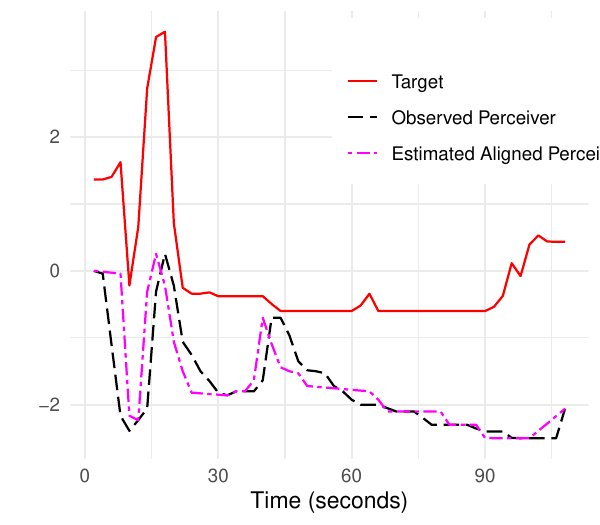}
    \includegraphics[width=0.42\linewidth]{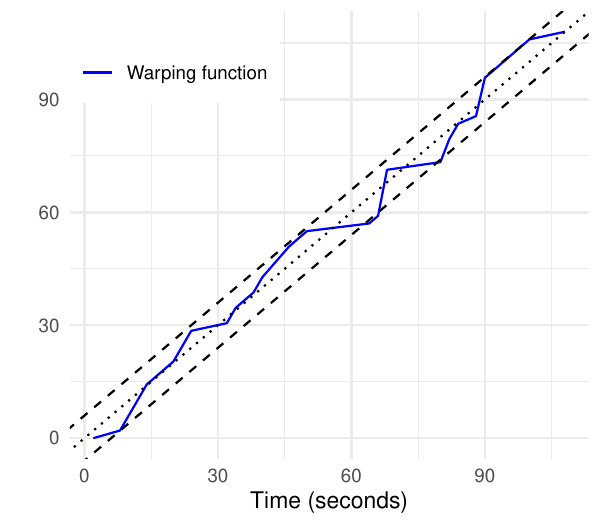}
    \caption{Penalized alignment}
    \label{subfig:DelvinFnpen}
    \end{subfigure}
    \caption{Example target and perceiver's emotion ratings of \cite{devlin2014not}. (Left): target $x$ (solid), perceiver's observed response $y$ (dash), and estimated perceiver's response $\hat{y} = y \circ \hat{\gamma}$ (dot dash). (Right): estimated warping function $\hat{\gamma}$.}
    \label{fig:Delvinexample}
\end{figure}
\subsection{Unpenalized SRVF leads to Over-Alignment}
\label{subsec:overalignment}
While the SRVF representation leads to several theoretical benefits, one main disadvantage of the unpenalized SRVF for studying EA is that the estimated perceiver function $y\circ\hat{\gamma}_{u}(t)$  may be overaligned with the target $x$ and thus could differ from the perceiver’s latent function $a$. In other words, the unpenalized SRVF not only corrects for the Misalignment C in Figure \ref{fig:enter-label}, but also potentially removes inherent temporal Disagreement B.
%Particularly, if observed perceiver functions are noisy or include false peaks or valleys due to random variation, they may be incorrectly aligned with real peaks or valleys of targets. 

Figure~\ref{subfig:DelvinFnunpen} shows one example from the study in \cite{devlin2014not} demonstrating the result of the previous SRVF alignment obtained from (\ref{eq:unpenwarpsrvf}). %The left panel illustrates the target's $x(t)$, the perceiver's $y(t)$, and the estimated perceiver's rating sequence under the unpenalized alignment $y\circ\hat{\gamma}_{u}(t)$. 
In this study, the continuous ratings were recorded for 108 seconds and averaged over 2-second epochs. 
%Here, time $t$ is rescaled from 0 to 1. 
The alignment is obtained by using their SRVF representations $q_x$, $q_y$, and $(q_y, \hat\gamma_{u})$. The estimated \revision{inverse} warping function $\hat{\gamma}_{u}$ is plotted in the right panel of Figure~\ref{subfig:DelvinFnunpen}. When $\hat{\gamma}_{u}$ appears above the 45-degree line, it implies that the perceiver's response is delayed compared to the target, %and $\hat{\gamma}_{u}$ adjusts it accordingly, 
whereas $\hat{\gamma}_{u}$ below the 45 degree line indicates that the perceiver's response precedes the target. While it seems reasonable to expect that the perceiver’s perception of a particular emotion would lag behind the target’s actual expression of that emotion, there is evidence to suggest that people can make anticipatory perceptual judgements, especially when the stimuli are continuous and dynamic. For example, Thornton and Tamir \cite{Thornton2017} found that perceivers attend to emotion regularities and can predict up to two emotional transitions into the future. Koster-Hale and Saxe \cite{Koster-Hale2013} 
argued that the brain actively generates expectations about others’ emotions, thoughts, and behaviors (so not just passively reacting to them). They refer to this as “predictive coding.” In Figure~\ref{subfig:DelvinFnunpen}, the peak of the perceiver's response around $t=45$ seconds is considered as a response preceding the target's self-rating around $t=65$ seconds, and it is aligned accordingly by the unpenalized SVRF method.

%The above example clearly reveals the over-alignment problem of the unpenalized SVRF alignment method. . 

Therefore, from the left plot of Figure~\ref{subfig:DelvinFnunpen}, the unpenalized SRVF method misaligns the peak of the perceiver's response, occurring at approximately $t=40$ seconds, with the target's small peak at around $t=65$ seconds, which is likely just noise. This alignment suggests an improbable scenario where the perceiver predicts the target's emotional change 25 seconds in advance.  Psychological research has consistently shown that reaction time-delay is limited to a few seconds: 0.5 to 4 seconds \citep{levenson1988emotion}, 3 to 6 seconds \citep{nicolle2012robust}, 2 to 11 seconds \citep{mariooryad2014correcting}, and 0.48 to 6.24 seconds \citep{ringeval2015prediction}.
By disregarding this inherent limitation, unpenalized SRVF alignment overestimates synchronization between rating sequences, potentially leading to an unrealistic shape of the estimated warping function that exceeds the human exception bounds, and hence biased estimations of perceiver EA levels.

\section{Method}
\label{sec:method}
\subsection{Penalized Elastic Functional Alignment}
\label{subsection:pen}
Penalized alignment has been proposed to control the amount of alignment 
\citep{wu2011information,mitchell2022object,guo2022data} 
%Wu: neural spike train data analysis 
%Mitchell: geological surface satellite data analysis 
%Guo: landmark alignment
or to achieve smooth alignment \citep{srivastava2016functional}. To address the over-alignment issue inherent in the unpenalized SRVF method, an existing solution is to employ a penalized alignment approach by %that limits the maximum allowable shift.
%This is achieved 
by incorporating a penalty term into the unpenalized alignment optimization function (\ref{eq:unpenwarpsrvf}). This results in the following objective function:
\begin{eqnarray}
\Vert q_x-(q_y, \gamma) \Vert _2^2 + \lambda \mathcal{R}(\gamma),
\label{eq:penwarpsrvf}
\end{eqnarray}
where $\gamma$ is the \revision{inverse} warping function, $\lambda>0$ is a penalty parameter, and $\mathcal{R}(\gamma)$ is a penalty function.
Several penalty functions have been suggested in the literature, 
such as $\mathcal{R}(\gamma)= \| \sqrt{{\gamma}^
\prime}-\textbf{1}\|_2^2$ and $\mathcal{R}(\gamma)= \cos^{-1}(\langle \sqrt{{\gamma}^\prime},\textbf{1} \rangle)$, which are used to measure the differences between the SRVFs of $\gamma$ and the identity warping $\psi_{id}(t)=\gamma_{id}(t)=t$ by the squared $\mathbb{L}^2$ norm and the arc length, respectively, where $\textbf{1}$ is the constant function with value 1 and $\langle \cdot,\cdot\rangle$ denotes an inner product operator \citep{srivastava2016functional}.
%, or the arc length between SRVFs $\mathcal{R}(\gamma)= \cos^{-1}(\langle \sqrt{{\gamma}^\prime},\textbf{1} \rangle)$}, where $\langle \cdot,\cdot\rangle$ is an inner product \citep{srivastava2016functional}.%the second-order roughness penalty $\mathcal{R}(\gamma)=\int_0^1(\ddot{\gamma}^2(t)) dt$ for a smooth warping and 

The aforementioned penalty functions are inappropriate for current EA research. Primarily, it is challenging to select an optimal data-driven tuning parameter $\lambda$.  Common cross-validation procedures that split the data into independent training and test sets do not preserve the geometric features of the data.   % Given the unknown and impractical nature of determining the optimal alignment amount from  EA test data, selecting an appropriate $\lambda$ is problematic.  
Secondly, as reviewed in Section \ref{subsec:overalignment}, psychological research indicates that misalignment in perceiver ratings occurs within a specific temporal window of a few seconds. Existing penalty functions, however, focus on controlling the overall amount of warping, which does not directly translate to constraining alignment at each individual time point as required for EA studies.

To address these limitations, we introduce a novel penalized alignment method that directly incorporates the established temporal boundary for maximum perceiver misalignment as a penalty term. Specifically, we construct the optimal \revision{inverse} penalized warping function 
\begin{align}
\hat{\gamma}_p = \arginf_{\gamma \in \revision{\Gamma_{I}}} \Vert q_x-(q_y, \gamma) \Vert_2^2, \nonumber\\
\text{s.t. } \sup_t| \gamma(t)-\gamma_{id}(t)|\leq \nu,
\label{eq:optproblem}
\end{align}
where %$\gamma_{id}(t) = t$ is the identity warping and 
$\nu$ is the predefined upper limit of warping functions, corresponding to the maximum delay or advance observed in the perceivers' responses. Although the supremum norm limit $\nu$ plays the role of a tuning parameter, in practice, we often have prior knowledge about its value based on the research context, unlike the tuning parameter $\lambda$ in the existing approach \eqref{eq:penwarpsrvf}. Nevertheless, it is useful to perform a sensitivity analysis of the proposed method over a reasonable range of $\nu$.
We denote $\hat{\gamma}_p$ as the estimated \revision{inverse} warping function of penalized alignment, where the subscript ${p}$ stands for ``penalized.'' As $\nu \rightarrow 0$, $\hat\gamma_p \rightarrow \gamma_{id}$, so that no warping is allowed. On the other hand, if $\nu \geq \sup| \hat{\gamma}_{u
}-\gamma_{id}|$, the constraint in  \eqref{eq:optproblem} is inactive, then $\hat\gamma_p = \hat\gamma_{u}$. Consequently, any $\nu$ smaller than $\sup| \hat{\gamma}_{u}-\gamma_{id}|$ induces a shrinkage effect, pulling the unpenalized warping towards the identity warping function, akin to penalized regression. This interpretable penalty mechanism enables our proposed penalized alignment to mitigate the risk of over-alignment, resulting in more plausible warping estimates and aligned responses. 

We note that under the constraint \eqref{eq:optproblem}, the Fisher-Rao phase distance $d_p$ defined by (\ref{eq:fisherrao}) is still valid to measure the difference between the phase of two functions, with the exception that $\hat\gamma_{u}$ is replaced by $\hat\gamma_p$. The proof of Lemma~\ref{lemma:1} is given in Section~S1 of the Supplementary Material.
\begin{lemma}
\label{lemma:1}
The Fisher-Rao phase distance between $x$ and $y$
is estimated by
$d_p(x,y) = \cos^{-1}\left(\int_0^1 (\sqrt{{\hat{\gamma}^\prime}_{p}(t)} dt\right)$ %is the Fisher-Rao phase distance for \eqref{eq:optproblem}.
\end{lemma}

\subsection{Computing the penalized SRVF alignment}
A discrete approximation for the solution of the optimization problem specified in \eqref{eq:optproblem} can be found by using the following dynamic programming algorithm (DPA) \citep{srivastava2016functional}. Assume both the SRVF functions of the target and the perceiver $q_x$ and $q_y$ are observed at $T+1$ time points, $t_0 < t_1 < t_2 < \cdots \leq t_T$. Without loss of generality, we assume that $t_0 = 0$ and $t_T = 1$, and that these time points are equally spaced, i.e., $t_m = m/T$ for $m=0,\ldots, T$.
The \revision{inverse} warping function $\gamma$ matches the point $(q_y, \gamma)$ with the point $q_x$, so $\gamma$ can be viewed as a graph with a collection of points $(t_m, \gamma(t_m))$, from $(0, 0)$ to $(1,1)$ in $\mathbb{R}^2$. We assume that within each interval $[t_m, t_{m+1}]$, the function $\gamma(t)$ can be approximated by a straight line, so the final estimate for $\hat\gamma$ is a piecewise linear path.  Since $\gamma$ is non-decreasing, the slope of this graph is always strictly between 0 and 90 degrees.  Furthermore, the cost function in \eqref{eq:optproblem} can be approximated by
\begin{equation}
\int_{0}^{1} \left\{q_x(t) - q_y\left(\gamma(t)\right)  \sqrt{{\gamma}^\prime(t)}\right\}^2 dt \approx
\sum_{m=0}^{T} \int_{t_m}^{t_{m+1}} \left\{q_x(t) - q_y\left(\gamma_m(t)\right)  \sqrt{{\gamma}_m^\prime(t)}\right\}^2 dt,
\label{eq:2}
\end{equation}
where $\gamma_m (t)$ is a straight line connecting $(t_m, \gamma(t_m))$ and $(t_{m+1}, \gamma(t_{m+1}))$. The function on the right-hand side of \eqref{eq:2} is
additive over the graph, and hence enables the use of DPA. Our goal then is to find an optimal linear piecewise path from $(0, 0)$ to $(1,1)$ in $\mathbb{R}^2$ that minimizes \eqref{eq:2}, subject to the constraint that $\sup_{t \in \left\{t_1, \ldots, t_T\right\}} \vert \gamma(t) - t \vert \leq \nu$. Using DPA, we can construct this path recursively as follows. 

Given a feasible point $(t_k, t_l)$, i.e., $\vert t_l - t_k \vert \leq \nu$ in the graph,  let $\mathcal{N}_{k, l} = \{(k^\prime, l^\prime) \mid 0 \leq k^\prime < k, ~0 \leq l^\prime < l, \vert k^\prime - l^\prime \vert \leq \nu \} $ denote the set of all nodes in the graph that are allowed to go to $(t_k, t_l)$ by a straight line. Starting from $(0,0)$, if we have already determined and stored the cost of reaching nodes in $\mathcal{N}_{k, l}$, then the cost of reaching $(t_k, t_l)$ is given by
\begin{equation}
H_{k, l} = \min_{(k^\prime, l^\prime ) \in \mathcal{N}_{k, l}} \left(H_{k^\prime, l^\prime}  +  \int_{t_k}^{t_{k^\prime}} \left\{q_x(t) - q_y\left(\gamma(t)\right)  \sqrt{{\gamma}^\prime(t)}\right\}^2 dt \right),
\label{eq:3}
\end{equation}
where we initialize $H_{0,0} = 0$ and $H_{0, l} = H_{k, 0} =  \infty$ for any $l \neq 0$ and $k \neq 0$.
Let $(\hat{k}, \hat{l})$ be the nodes that minimize the right-hand side of \eqref{eq:3} and repeat the process for every possible point $(t_k, t_l)$ in the graph. Then the optimal curve $\hat\gamma_p$ is obtained by connecting all such points using piecewise linear curves. Note that compared to the standard DPA algorithm to align the two functions \citep{srivastava2016functional}, %\citet[Section 4.7]{srivastava2016functional}, 
we have modified the set of permitted nodes to account for the constraint imposed on the warping function. 

The algorithm is summarized in Algorithm~\ref{alg:cap}. 
\begin{algorithm}

\caption{SRVF alignments with sup constraints}\label{alg:cap}
\begin{algorithmic}
\State Set $\mathcal{N}_{k, \ell} =  \{(k^\prime, l^\prime) \mid 0 \leq k^\prime < k, ~0 \leq l^\prime < l, \vert k^\prime - l^\prime \vert \leq \nu \}$
\State Set  $H_{0, 0} = 0$, $H_{0, l} = H_{k, 0} =  \infty$ for any $l \neq 0$ and $k \neq 0$.
\For{ $ (0, 0 < (t_k, t_l) < (1,1)$}
    \State Find ($k^\prime, l^\prime$) that minimizes the right-hand side of \eqref{eq:3}
    \State Compute $H_{k, l}$
\EndFor 
\State Initialize $(t_k, t_l) = (1,1)$ \Comment{Construction of $\hat\gamma_p$}
\While{$(t_k, t_l) \neq (0, 0)$}
    \State Draw a straight line between $(t_k, t_l)$ and $(t_{k^\prime}, t_{l^\prime})$
    \State Set $(t_k, t_l) = (t_{k^\prime}, t_{l^\prime})$
\EndWhile
\end{algorithmic}
\end{algorithm}
\FloatBarrier

%\textcolor{blue}{Jing: Linh, I think we can add a graph using our data like the following example to demonstrate the idea of time warping of sequences.}

Because the temporal window can vary according to the emotions and the modality \citep{gunes2010automatic,gunes2013categorical}, we used three thresholds of $\nu=6, 8$, and $10$ seconds for EA applications in Section~\ref{sec:data_app}. Figure~\ref{subfig:DelvinFnpen} shows penalized alignment of the same example presented in Figure~\ref{subfig:DelvinFnunpen}, using an upper limit of warping function differences of six seconds ($\nu=6$). Since $\sup \vert \hat\gamma_{u} - \gamma_{id} \vert = 23.39 > \nu=6$ for unpenalized alignment, penalized alignment shrinks the estimated \revision{inverse} warping function $\hat{\gamma}_p$ toward the identity warping function. Consequently, the resulting \revision{estimated perceiver latent function $\hat{y} = y \circ \hat{\gamma}_p$} does not exhibit peaks or valleys that deviate from the observed perceiver sequence by more than six seconds.

\section{Simulation Study}
\label{sec:simulation}

To demonstrate the performance of our functional alignment approach, we conducted a number of simulation studies. It is challenging to use real EA data to evaluate functional alignment methods because perceivers' latent ratings are unknown. However, we generated perceivers' latent responses from the real target ratings %without phase misalignment, denoted as perceivers' aligned responses in simulations, 
and used them to compare different alignment methods.

\subsection{Simulation 1}
\label{subsec:sim1}

In this simulation, we evaluated the effectiveness of various alignment methods. %We followed the idea of \citet{matuk2022bayesian} to generate the functional data. Their primary idea is, that since we decompose a function into its phase and amplitude components, we can model the two components separately.
To approximate the settings in real--data applications, we used the four targets $x_j(t)$ directly derived from the real target data of \citet{devlin2014not} corresponding to four videos: high intensity positive, low intensity positive, high intensity negative, and low intensity negative, for $j=1,\ldots, 4$.  We smoothed these raw data using the cubic smoothing spline and recorded the functional values for $300$ evenly-spaced time points ($t=0,1,\ldots,299$).

Next, we generated $n=500$  perceivers' latent responses for each target function $x_j(t)$ using the following model
\[
a_{ij}(t) = \epsilon_{ij}(t)\, x_j(t) + u_{ij}(t),
\]
for $i = 1, \ldots, n $. Here, we set $\epsilon_{ij}(t) = K_{h}(W_{ij}(t) + 1)$, where $W_{ij}(t)$ is a one-dimensional Wiener process (i.e., Brownian motion) at time $t$~\citep{morters2010brownian},  $u_{ij}(t) = K_{h}(S_{ij}(t))$ with $S_{ij}(t)$ being a standardized random walk at time $t$ that is obtained by cumulatively summing the standard normal $N(0, 1)$ noise and applying a standardization transformation. We denote $K_h$ as the Gaussian kernel smoothing with bandwidth $h$, and in this simulation, we used $h=20$ to ensure both $\epsilon_{ij}$ and $u_{ij}$ are smooth.
%\begin{equation}
%    a_i(t) = Q^{-1}\left(\beta_0(t) + \beta_1(t) q_x(t)\right) + \varepsilon_i^a(t),
%    \label{eq:sim_a}
%\end{equation}
%where $Q^{-1}(q_f)$ denotes the inverse transformation from an SRVF $q_f$ to the original function $f$ and $q_x$ is the SRVF of the target $x$. We set $\beta_0(t) = 8\sin(\pi t/50)$ and $\beta_1(t) = 5\sin(\pi t/100)$, followed by the setting of \citet{ghosal2020variable}, and $\varepsilon^a_i \sim N(0,0.1)$.
Then we generated the  perceiver's observed response $y_{ij}$ using the perceiver's latent rating $a_{ij}$ by
\begin{equation}
    y_{ij}(t) = (a_{ij}\circ \psi_{ij})(t), %+ \varepsilon_i^y(t),
    \label{eq:sim_y}
\end{equation}
where $\psi_{ij} \in \{ \psi \mid \psi \in \Gamma_I,  \sup| \psi(t)-t| = \eta_{ij}$ for $t\in[0,299] \}$ is the warping function and $\eta_{ij}$ is the true individual upper limit of the warping amount. We first generated the warping functions randomly using the  \textit{rgam} function in the R package \texttt{fdasrvf}
\cite{tucker2023fdasrvf}, and then rescaled them such that $\sup| \psi_{ij}(t)-t| = \eta_{ij}$. %and $\varepsilon^y_i \sim N(0,0.1)$. 
%Here, each true latent perceiver of $a_{ij}$ is warped by $\psi_{ij} \in \{ \psi \mid \psi \in \Gamma_I,  \sup| \psi(t)-t| = \eta_{i}$ for $t\in[0,299] \}$, where $\eta_{i}$ is the true individual upper limit of the warping amount. 
With that simulation configuration, the true correlation between $a_{ij}$ and the target $x_j(t)$ has the mean around \{0.65, 0.66, 0.67, 0.66\} with standard deviation \{0.24, 0.27, 0.23, 0.25\} for all $j=1,\ldots, 4$, respectively.
%

%The alignment methods aim to recover the perceiver's latent response by aligning the observed perceiver response $y_{i}$ to the target $x$. Let $\hat{y}_{i}(t)=y_{i}\circ \hat{\gamma}_{i}(t)$ denote the estimated perceiver response. The optimal alignment is expected find $\hat{y}$ close to $a_i$ and the optimal warping function $\gamma_i^*$ will the inverse of $\psi_{i}^{-1}(t)$. 

We considered five different methods to align the observed perceiver response to the target, including (1) no alignment, (2) optimal fixed delay, (3) unpenalized SRVF alignment, (4) the squared $\mathbb{L}^2$ norm penalty SRVF alignment, and (5) our proposed penalized SRVF alignment. Let $\hat{y}_{ij}(t)=y_{ij}\circ \hat{\gamma}_{ij}(t)$ denote the estimated perceiver's response, where $\hat{\gamma}_{ij}(t)$ denotes an estimated \revision{inverse} warping function from one of the above methods. The no alignment option assumes the identity inverse warping $\hat{\gamma}_{ij}(t)=t$. For the optimal fixed delay method (2), we found the optimum amount of delay $0\leq\delta\leq \nu$ that achieves the smallest $\mathbb{L}^2$ distance between $q_{x_j}$ and $(q_{y_{ij}},\hat{\gamma}_{ij})$, where $\hat{\gamma}_{ij}(t)=0$ if $t=0$, $\hat{\gamma}_{ij}(t)=t+\delta$ if $0<t<1-\delta$, and $\hat{\gamma}_{ij}(t)=1$ otherwise. The \revision{inverse} warping functions of the unpenalized and penalized SRVF alignments were obtained by solving \eqref{eq:unpenwarpsrvf} and \eqref{eq:optproblem}, respectively. The squared $\mathbb{L}^2$ norm penalty SRVF alignment implements the penalty $\hat{\gamma}_{ij} = \arginf_{\gamma \in \revision{\Gamma_{I}}} \Vert q_{x_j}-(q_{y_{ij}}, \gamma) \Vert_2^2 + \lambda \Vert \sqrt{\gamma^\prime(t)} - \textbf{1} \Vert^2_\revision{2}$ where $\textbf{1}$ is the constant function with value one \citep{srivastava2016functional}. To the best of our knowledge,  an optimal method for selecting $\lambda$ has not been established in the literature, so %and cross-validation is not applicable, 
we implemented the method with $\lambda = 0.01$. We leave the investigation of optimal selection strategies for 
$\lambda$ to future research.  %We note that the result for this method can be sensitive to our choice of the tuning parameter $\lambda$, and how to select its optimal value is out of scope for this paper.

In the simulation, we set the alignment warping limit for the penalized SRVF to $\nu\in \{6,8,10\}$ seconds regardless of the true warping limit $\eta_{i}$ to reflect the real-world cases where the true warping limit is unknown. Also, to account for individual warping variations, we examined two different settings of $\eta_{ij}$, including a constant $\eta_{ij} = \nu$ seconds for all $i=1,\ldots,n$ and $j=1,\ldots, 4$, and a varying $\eta_{ij}$ randomly generated from a Gamma distribution $\Gamma(k,\theta)$ with $k = \nu$ being the shape and $\theta=1$ being the scale parameter of the Gamma distribution. %We also conducted simulation studies with constant The simulation results of $\nu=6$ and $\nu=10$ seconds are reported in Tables~?? in Supplementary material.

We evaluated the performance of the alignment methods with two metrics. 
%First, we compare the correlations between the perceiver's latent response and the estimated perceiver's response $\rho_{a}=\rho(a_{i},\hat{y}_{i})$. Under the optimal alignment where $\hat{\gamma}_{i}(t) \approx \psi^{-1}_{i}$, the correlation is expected to be close to 1. 
First, we computed the average $\mathbb{L}^2$ distance between the  perceiver's latent function $a_{ij}$ and the estimated functions $\hat{y}_{ij}$ by $d_a=\| \hat{y}_{ij} - a_{ij}\|_2^2$. The closer $d_a$ gets to zero, the more accurate estimation of the perceiver's latent response.  
%Third, we evaluate the differences between the SRVF representations of the perceiver's aligned and estimated responses, $d^{q}_u=\Vert q_{a_{i,j}}-q_{\hat{y}_{i,j}}\Vert_2^2$. A small value of $d^{q}_u$ suggests that the amplitude variability between the perceiver's aligned and estimated responses is small. 
Second, we computed the average bias between the true and the estimated correlations to the target, $n^{-1}\sum_{i=1}^{n}\left\{\rho(x_j,\hat{y}_{ij})-\rho(x_j,a_{ij})\right\}$. Here, $\rho(x_j,\hat{y}_{ij})$ is a commonly used metric for measuring EA, and $\rho(x_{j},a_{ij})$ can be considered as the true correlation that the alignment methods aim to achieve. We reported the results for each target separately.    %Due to the small scale of the bias, we report values multiplied by 100.
%Second, we computed the mean error (ME) and the mean absolute error (MAE) between the true and the estimated correlations to the target, $\text{ME}=(I)^{-1}\sum_{i=1}^I\left\{\rho(x_i,\hat{y}_{i})-\rho(x_i,a_{i})\right\}$ and $\text{MAE}=(I)^{-1}\sum_{i=1}^I\vert\rho(x_i,\hat{y}_{i})-\rho(x_i,a_{i})\vert$. Here, $\rho(x_i,\hat{y}_{i})$ is a commonly used metric for measuring EA, and $\rho(x_i,a_{i})$ can be considered as the ideal correlation that the alignment methods aim to achieve.

\tabcolsep 0.2em
\begin{table}[th]
\centering
\caption{
    Performance of different alignment methods in the simulation studies under different warping limits $\eta$, $d_a$ between the estimated perceiver $\hat{y}$ and the true latent perceiver $a$, and the $(10\times)$ bias of the estimated correlation between the true latent perceiver and the target.  The lowest absolute bias and the lowest $d_a$ are highlighted for each row. Standard errors are included in the parentheses. }
    \label{tab:sim_summary_parameters}
\begin{tabular}[t]{lll>{}ccccc}
\toprule
$\eta$ & Target &  Metric & Pen. SRVF & $\mathbb{L}^2$ SRVF & Unpen. SRVF & Opt. Fixed & No Alignment\\
\midrule
8 & \multirow{2}{1cm}{High Neg} & $d_a$ & \textbf{4.81 (1.36)} & 6.74 (3.69) & 11.76 (5.1) & 6.71 (3.54) & 5.62 (1.46)\\
 &  & Bias & \textbf{0.02 (0.39)} & 0.67 (0.91) & 1.37 (1.53) & -0.36 (1.12) & -0.17 (0.41)\\
\addlinespace
 & \multirow{2}{1cm}{Low Neg} & $d_a$ & {8.33 (5.12)} & \bfseries 4.23 (2.44) & 9.22 (4.04) & 9.03 (5.23) & 10.56 (6.1)\\
 &  & Bias & \textbf{-0.68 (1.79)} & 0.69 (1.17) & 1.47 (1.75) & -0.82 (2.14) & -1.55 (2.05)\\
\addlinespace
 & \multirow{2}{1cm}{High Pos} & $d_a$ & \textbf{5.78 (1.62)} & 7.17 (3.88) & 11.59 (5.24) & 7.88 (3.35) & 6.04 (1.72)\\
 &  & Bias & \textbf{-0.01 (0.87)} & 0.63 (1.44) & 0.71 (2.25) & -0.21 (1.74) & -0.13 (0.88)\\
\addlinespace
 & \multirow{2}{1cm}{Low Pos} & $d_a$ & {5.92 (1.86)} & \bfseries 4.84 (3.23) & 8.3 (3.52) & 7.82 (3.83) & 6.59 (1.86)\\
 &  & Bias & \textbf{-0.13 (0.61)} & 0.55 (0.84) & 0.67 (1.1) & -0.76 (1.59) & -0.31 (0.65)\\
\addlinespace[1em]
$\Gamma(8,1)$ &  \multirow{2}{1cm}{High Neg} & $d_a$ & \textbf{4.75 (1.82)} & 6.64 (3.56) & 11.30 (5.03) & 6.57 (3.72) & 5.52 (1.96)\\
 &  & Bias & \textbf{0.05 (0.43)} & 0.72 (0.92) & 1.27 (1.44) & -0.28 (1.03) & -0.14 (0.45)\\
\addlinespace
 &  \multirow{2}{1cm}{Low Pos} & $d_a$ & {8.02 (4.98)} & \bfseries 4.24 (2.60) & 9.55 (4.24) & 9.26 (5.51) & 10.49 (5.95)\\
 &  & Bias & \textbf{-0.6 (1.73)} & 0.68 (1.27) & 1.42 (1.78) & -0.86 (2.27) & -1.46 (1.98)\\
\addlinespace
 & \multirow{2}{1cm}{High Pos}  & $d_a$ & \textbf{5.69 (1.93)} & 7.16 (3.91) & 12.18 (5.81) & 7.7 (3.41) & 5.94 (2.00)\\
 &  & Bias & \textbf{-0.09 (0.83)} & 0.59 (1.26) & 0.55 (2.52) & -0.32 (1.62) & -0.20 (0.84)\\
\addlinespace
 & \multirow{2}{1cm}{Low Pos}  & $d_a$ & {5.98 (2.20)} & \bfseries 4.98 (3.17) & 8.40 (3.53) & 7.96 (4.17) & 6.57 (2.08)\\
 &  & Bias & \textbf{-0.11 (0.73)} & 0.68 (0.98) & 0.77 (1.31) & -0.66 (1.98) & -0.29 (0.72)\\
\bottomrule
\end{tabular}
\end{table}

Table \ref{tab:sim_summary_parameters} shows the performance metrics for the case $\eta_{ij} = 8$ and $\eta_{ij} \sim \Gamma(8,1)$. Results from additional settings, which lead to similar conclusions, are provided in Section S2 of the Supplementary Materials.  Among the five alignment methods evaluated, \revision{the proposed penalized SRVF approach consistently outperforms the others in producing the least biased estimation of EA in all the considered simulation designs.} In addition,
the average amplitude distance $d_a$ of the proposed penalized SRVF is the smallest for the high negative and high positive targets. For the low negative and low positive targets, the $\mathbb{L}^2$ penalized SRVF shows the lowest average $d_a$ but yields much larger bias. Considering both metrics, the results imply that the proposed method makes the most accurate estimation of the phase shift. 
%Third, the average SRVF distance between the perceiver's true and aligned responses of the penalized SRVF is the smallest. This suggests that the penalized SRVF makes the least amplitude variability after alignment. 

The unpenalized SRVF, optimal fixed, and no alignment methods all yield less accurate estimates of $a_{ij}$ compared to the proposed penalized SRVF.
Among them, the unpenalized SRVF produces the largest value of $d_a$, primarily because it tends to over-align the perceiver’s observed response $y_{ij}$ to the target’s rating $x_j$, leading to distorted estimates $\hat{y}_{ij}$. Additionally, the optimal fixed alignment method often results in the highest standard errors and bias in $d_a$, indicating that it provides inconsistent estimates of the perceiver’s latent ratings.

\subsection{Simulation 2}

In practical applications, the true warping limit $\eta$ is typically unknown. As a result, the alignment warping limit $\nu$  must be chosen based on approximate prior knowledge, which may not perfectly match the true value. To examine the impact of this mismatch, we conducted a simulation study using the same data generation process described in Section~\ref{subsec:sim1}. Specifically, we evaluated 21 different true warping limits $\eta \in \{0,0.8,\ldots,8,\ldots,15.2, 16\}$ seconds, while fixing the alignment warping limit at $\nu=8$ seconds. %We apply the four alignment methods to align the observed perceivers' responses. 

\begin{figure}[!ht]
    \centering
    \begin{subfigure}{0.48\linewidth}
    \centering
    \includegraphics[width=0.9\linewidth]{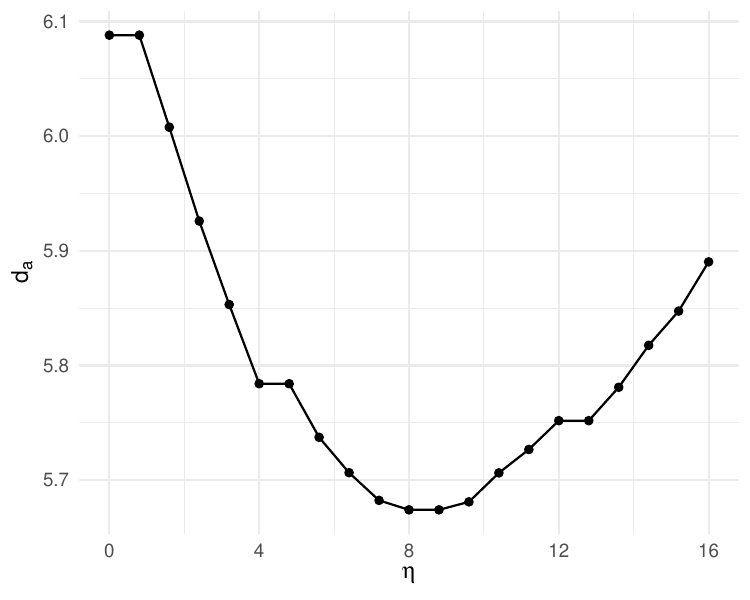}
    \caption{$d_a$}
    \label{subfig:sim2_da}
    \end{subfigure}
    \begin{subfigure}{0.48\linewidth}
    \centering
    \includegraphics[width=0.9\linewidth]{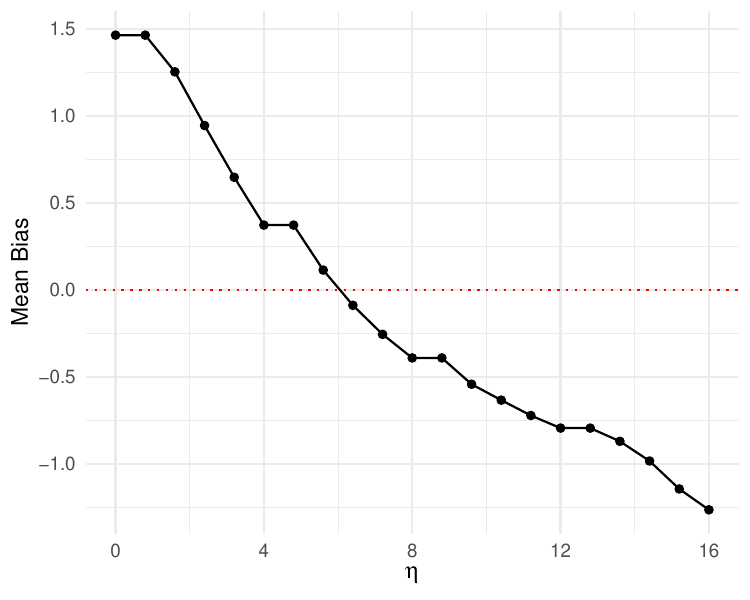}
    \caption{Mean Bias}
    \label{subfig:sim2_mse}
    \end{subfigure}
    \caption{The penalized SRVF alignment results under 21 different true warping limits $\eta \in \{0,0.8,\ldots,8,\ldots,15.2, 16\}$ seconds when the upper limit of alignment is $\nu=8$ seconds. The red dotted line in the mean bias plot marks the unbiased level.}
    \label{fig:sim2}
\end{figure}

Figure~\ref{fig:sim2} represents two performance metrics ($d_a$ and bias) of the penalized SRVF method across 21 different true warping limits $\eta$. The results illustrate the method’s robustness to variation in the true warping limit. Notably, when $\eta$ falls within approximately two seconds of the alignment limit $\nu = 8$ seconds,  the penalized SRVF method achieves low $d_a$ and minimal bias, indicating accurate alignment and estimation. These findings suggest that even without precise knowledge of the true warping limit, selecting $\nu$ within a reasonable range yields reliable performance, underscoring the method’s practical utility in real-world applications.
%Also, the results are closer to the case when the true warping limit is correctly identified ($\nu=\eta=30$). These findings indicate that selecting $\nu$ within a reasonable warping limit range yields robust alignment outcomes for the penalized SRVF method, even when the exact degree of true warping limit $\eta$ is unknown.

\subsection{Simulation 3}
In this simulation, we evaluated the performance of the alignment methods under different levels (i.e., high/medium/low) of EA. Perceivers' latent ratings $(i=1,\ldots,500)$ were generated following the idea proposed by \citet{matuk2022bayesian},
\begin{equation}
    a_i(t) = Q^{-1}\left(\beta_{0i}(t) + \beta_{1i}(t) q_x(t)\right) + \varepsilon_i^a(t),
    \label{eq:sim_a}
\end{equation}
where $Q^{-1}(q_f)$ denotes the inverse transformation from an SRVF $q_f$ to the original function $f$ and $q_x$ is the SRVF of the target rating for the low intensity positive video $x$. The parameters, $\beta_{0i}(t) = \alpha_{0i}\sin(\pi t/50)$ and $\beta_1(t) = \alpha_{1i}\sin(\pi t/100)$, were chosen based on the setting in \citet{ghosal2020variable}, with $\varepsilon^a_i \sim N(0,0.1^2)$. 

We generated perceivers' observed responses following the same data generation process as in~\eqref{eq:sim_y}, setting the alignment warping limit to $\nu=8$ and drawing the true sup-norm limit from a Gamma distribution, $\eta_i\sim \text{Gamma}(8, 1)$. To simulate varying levels of EA, we defined three conditions: for high EA, $\alpha_{0i},\alpha_{1i} \sim N(0,0.05^2)$; for medium EA, $\alpha_{0i},\alpha_{1i} \sim N(5,2^2)$; and for low EA, $\alpha_{0i} \sim N(0,0.5^2)$ and $\alpha_{1i} \sim N(0.1,0.1^2)$. The resulting average correlations between the target ratings and the perceivers’ latent ratings, $\rho(a_i,x)$, are approximately 0.81, 0.62, and 0.25 for the high, medium, and low EA conditions, respectively.

\begin{table}[!ht]
\caption{
    Comparison results based on \revision{$d_a$ between the estimated perceiver $\hat{y}$ and the true latent perceiver $a$ and the $(10\times)$ bias of the estimated correlation between the true latent perceiver and the target}. The best metrics are highlighted for each row. }
    \label{tab:sim_summary_EA}
\begin{tabular}{ll>{}lllll}
\toprule
EA & Metric & Pen. SRVF & $\mathbb{L}^2$ SRVF & Unpen. SRVF & Opt. Fixed & No Alignment\\
\midrule
 & $d_{a}$ & 3.33 (0.72) & \textbf{3.30 (0.69)} & 3.75 (0.58) & 4.39 (1.47)  & 3.57 (0.63) \\
\addlinespace[.2em]
\multirow[t]{-2}{*}
{\raggedright\arraybackslash High} & Bias & \textbf{0.04 (0.34)} & 0.50 (0.29)  & 0.83 (0.24) & -0.50 (0.70)  & -0.08 (0.33) \\
%{\raggedright\arraybackslash High} & Bias & \textbf{0.00 (0.03)} & 0.05 (0.03) & 0.08 (0.02) & -0.05 (0.07)  & -0.01 (0.03) \\
\addlinespace[.2em]
%\multirow[t]{-3}{*}{\raggedright\arraybackslash High} & MAE & \textbf{0.02 (0.02)} & 0.05 (0.03) & 0.08 (0.02)  & 0.06 (0.06) & 0.02 (0.02)\\
\addlinespace[1em]

 & $d_{a}$ & \textbf{5.10 (1.75)} & 5.77 (3.04) & 8.13 (3.11) & 6.72 (3.06)  & 5.33 (1.61) \\
\addlinespace[.2em]
\multirow[t]{-2}{*}
{\raggedright\arraybackslash Med} & Bias & \textbf{0.01 (0.56)} & 0.58 (0.84) & 1.43 (0.82) & -2.17 (2.17)  & -0.22 (0.66) \\
%{\raggedright\arraybackslash Med} & Bias & \textbf{0.00 (0.06)} & 0.06 (0.08) & 0.14 (0.08) & -0.22 (0.22)  & -0.02 (0.07) \\
\addlinespace[.2em]
%\multirow[t]{-3}{*}{\raggedright\arraybackslash Med} & MAE & \textbf{0.04 (0.04)} & 0.07 (0.08)  & 0.14 (0.08)  & 0.22 (0.21) & 0.05 (0.05)\\
\addlinespace[1em]

 & $d_{a}$ & \textbf{7.80 (2.62)} & 12.08 (5.55) & 14.25 (5.07) & 11.35 (5.11) & 7.82 (2.54) \\
\addlinespace[.2em]
\multirow[t]{-2}{*}
{\raggedright\arraybackslash Low} & Bias & 0.18 (0.62) & 2.06 (1.39) & 2.01 (0.87) & -0.11 (2.98)  & \textbf{-0.03 (0.65)} \\
%{\raggedright\arraybackslash Low} & Bias & 0.02 (0.06) & 0.21 (0.14) & 0.20 (0.09) & -0.01 (0.30)  & \textbf{-0.00 (0.07)} \\
\addlinespace[.2em]
%\multirow[t]{-3}{*}{\raggedright\arraybackslash Low} & MAE & \textbf{0.05 (0.04)} & 0.21 (0.13) & 0.20 (0.09)  & 0.25 (0.17) & 0.05 (0.04)\\
\bottomrule
\end{tabular}
\end{table}

Table~\ref{tab:sim_summary_EA} summarizes \revision{two} evaluation metrics for the five alignment methods across three levels of EA. The proposed penalized SRVF method demonstrates strong performance regardless of the EA levels. When EA is high, the $\mathbb{L}^2$ penalized SRVF achieves a slightly lower average $d_a$ than the penalized SRVF, but the latter yields the lowest average bias. \revision{The no alignment method also archives comparable low average bias to the penalized SRVF method.} At the medium EA level, the penalized SRVF outperforms all other methods. In the low EA condition, \revision{the penalized SRVF's $d_a$} is comparable to the no alignment approach \revision{but the no alignment method shows the lowest bias. The benefit of adjusting the observed rating is expected to be limited given the weak association between the target rating and the perceiver’s latent rating.} In contrast, the other alignment methods perform significantly worse than the penalized SRVF, particularly under low and medium EA conditions.

\FloatBarrier

\section{Data application}
\label{sec:data_app}
\subsection{Study on Social Empathy}
\label{sec:data_social}
In the first data application, we analyzed a dataset from \citet{devlin2014not}, which consists of 121 perceivers’ empathy responses of four distinct videos in which the targets discuss emotional events in their lives. 
%\textcolor{blue}{Chul: Linh, should we shorten the target explanation becaues we explained it in simulation study?}
The four videos vary in valence (positive or negative) and intensity (high or low), resulting in four heterogeneous videos, including high-positive, low-positive, high-negative, and low-negative. Participants provided continuous 9-point scale ratings of target emotions while watching each video. These perceiver ratings were compared to the target’s self-ratings. Following standard functional data analysis practices
\citep{srivastava2016functional}, 
%\citep[Section 4.2]{srivastava2016functional}, 
we preprocessed the data by smoothing target and perceiver ratings using cubic smoothing splines with the default setting of the \textit{smooth.spline} function in R and interpolating the estimated functions on a 300-point equidistant grid within the observed time interval. The goal of the subsequent analysis is to measure the level of EA for each perceiver, quantified by the correlation between the perceiver's latent ratings and the target's ratings.

Figure \ref{fig:delvin} clearly illustrates the misalignment between perceiver and target ratings. \citet{devlin2014not} did not account for this misalignment, measuring EA as a monotonic transformation of the Pearson correlation between the two rating sequences. We applied both unpenalized and penalized SRVF alignments, as these methods offer more flexible time warping than fixed delay alignment. Here, we present results for the penalized alignment with a threshold of $\nu=8$ seconds. Results for thresholds of 6 and 10 seconds are included in Section~S3.1 of the Supplementary Material.

To quantify the degree of warping, we computed the phase distance ($d_p$) between the estimated \revision{inverse} warping function under each alignment method and the identity warping function for each video. The summary statistics for this measure can be found in Table~S2 of the Supplementary Material. Figure \ref{fig:delvin_misalignments_pd} reveals that the unpenalized SRVF alignment consistently produces warping functions farther from the identity function than the penalized SRVF alignment, indicating the latter's effectiveness in reducing excessive warping, where the $p$-values (Table~S3 in the Supplementary Material) corresponding to the $t$-tests for the mean differences between penalized SRVF and other method are very close to zero.

\begin{figure}[!ht]
    \centering
    \includegraphics[width=\linewidth, page=3]{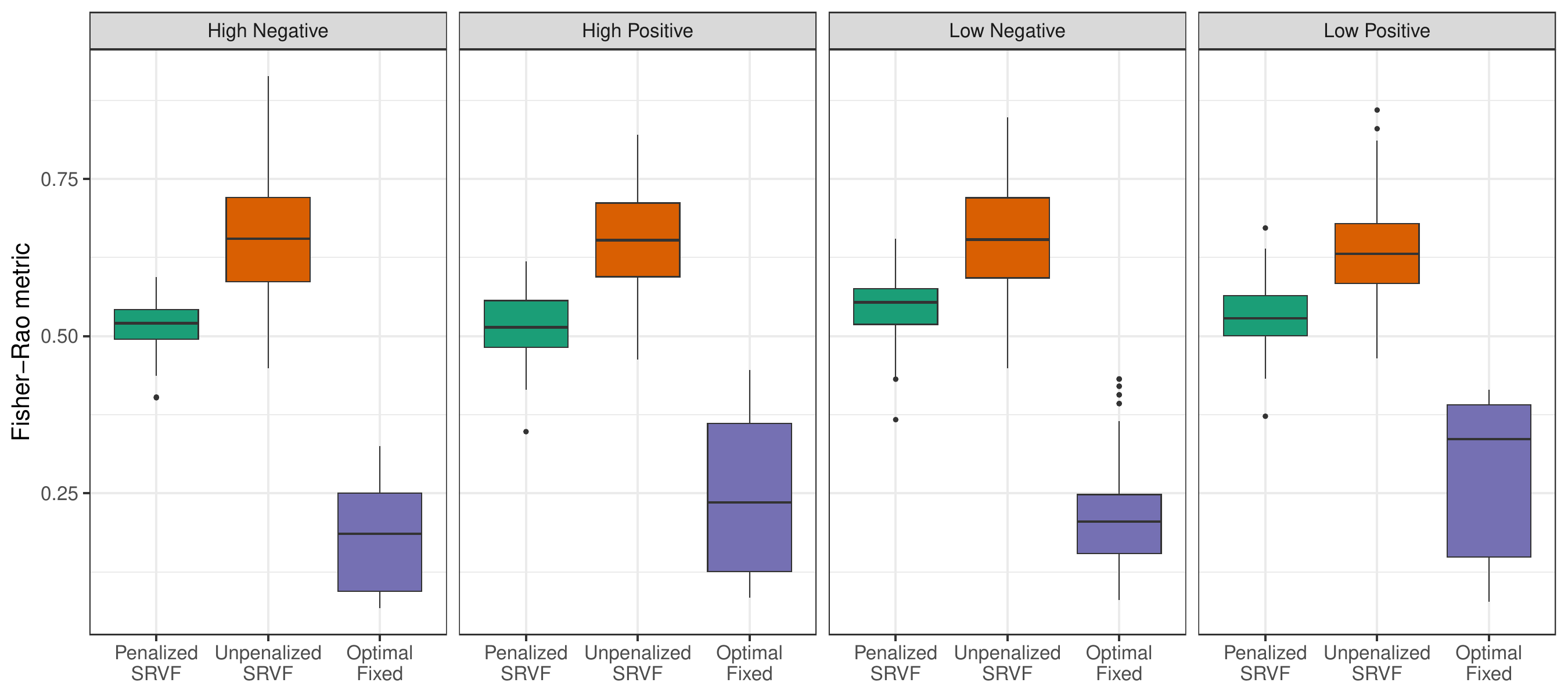}
    \caption{Boxplots for the estimated amount of warping, as measured by the Fisher-Rao metric between the identity warping $\gamma_{id}$ and the estimated warping function using unpenalized SRVF and penalized SRVF method, with $\nu = 8$ seconds for each video. }
    \label{fig:delvin_misalignments_pd}
\end{figure}

We subsequently calculated the Pearson correlation between each perceiver's aligned ratings and the target's ratings, which were used as the EA measure.  Unlike the simulation studies, the correlation coefficient $\rho(a,x)$ between the target ($x$) and the perceiver's latent response ($a$) is not observable because the perceiver's latent response is not known. Instead, we compared these EA measures to those obtained without alignment (identity warping), referred to as identity correlations. Notably, about 2\% of the cases exhibited negative correlations between perceiver and target ratings under identity warping. As a data pre-processing step, we removed those cases based on the concern that they may exhibit fundamentally different empathy patterns from the general population of perceivers.

\begin{figure}[!ht]
\begin{subfigure}{\textwidth}
    \centering
    \includegraphics[width=\linewidth, page=1]{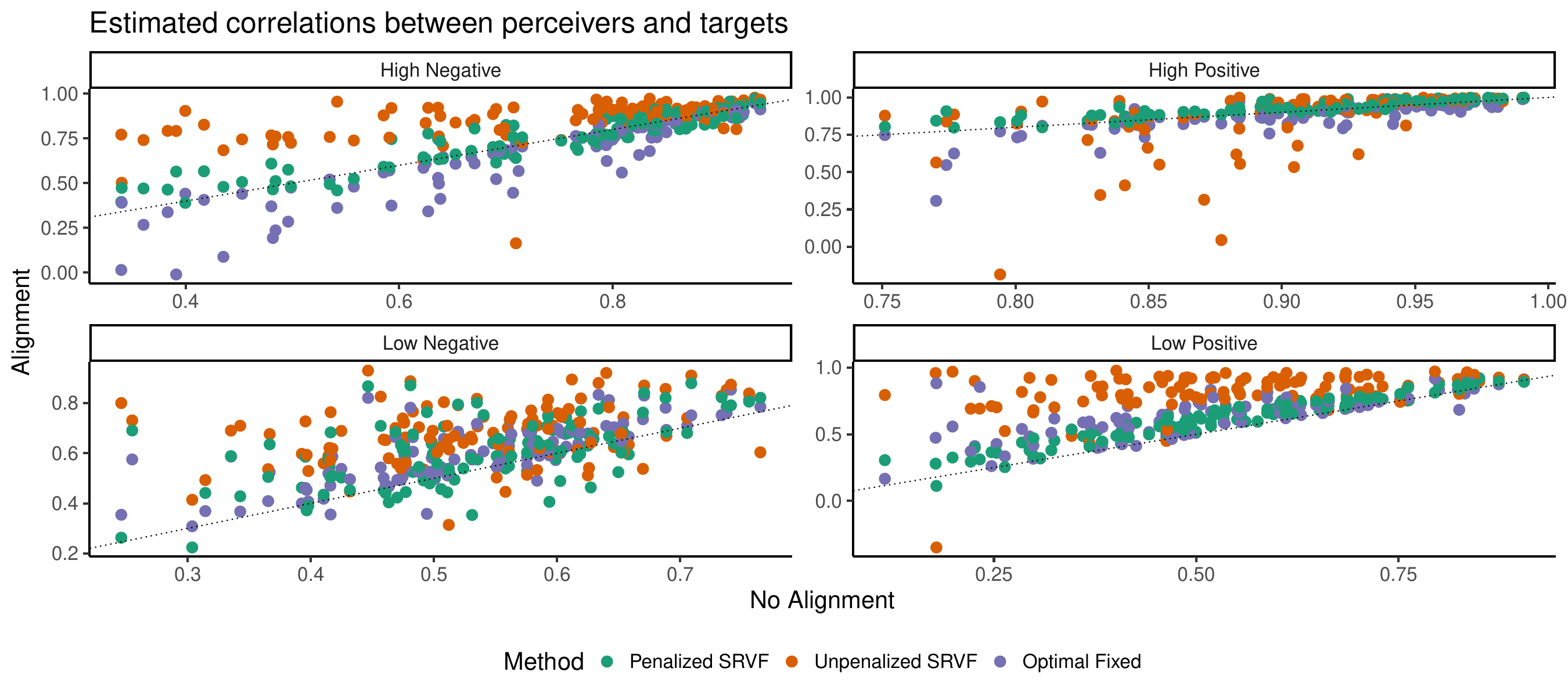}
    \caption{Scatterplots of correlation between target's ratings and perceivers' ratings for each video. In each plot, the horizontal axis represents the correlation without alignment, and the vertical axis represents the correlations under unpenalized SRVF (orange) and penalized SRVF with $\nu = 8$ seconds (green), respectively. The dashed line represents the 45-degree line. }
    \label{fig:delvin_misalignments_scatterplot}
\end{subfigure}
\begin{subfigure}{\textwidth}
 \centering
    \includegraphics[width=\linewidth, page=3]{plots/Delvin_emmeans_plots_rmoutliers_06012025_appendix.pdf}
      \caption{Pairwise confidence intervals for the mean difference between the estimated correlations obtained from unpenalized SRVF, optimal fixed, and no alignment and the proposed penalized SRVF method.}
          \label{fig:Delvin_pairwise_CI}
   
    \end{subfigure}

   \caption{Results for EA estimates in the social EA study.}
\end{figure}

Figure \ref{fig:delvin_misalignments_scatterplot} presents scatterplots comparing these EA measures between target and perceiver ratings for pre- and post-aligned data across the four video conditions. The majority of points reside above the 45-degree line, indicating that accounting for misalignment generally increases EA measures compared to unaligned analyses. However, the unpenalized SRVF alignment often inflates EA  considerably, as observed in the simulation results %reported in Section S2.2 in the Supplementary Material. 
This is most pronounced in the low intensity positive video group (bottom right panel of Figure \ref{fig:delvin_misalignments_scatterplot}), where many unpenalized EA approach one, implying near-perfect empathy for most perceivers, an unrealistic outcome given the video's low expressiveness. Conversely, for the high intensity positive video group (top right panel), some unpenalized EA  fell substantially below identity EA because excessive warping distorted overall function trends. 

\revision{The proposed penalized SRVF alignment provides a reasonable compromise between the identity EA and the unpenalized SRVF EA}. For low-expressivity videos (bottom row, Figure \ref{fig:delvin_misalignments_scatterplot}), penalized alignment EA measures generally exceed those from no alignment, likely due to increased misalignment challenges under reduced emotional cues. It is also interesting to find that for videos under positive emotion (second column, Figure~\ref{fig:delvin_misalignments_scatterplot}), the EA with the penalized SRVF alignment has a trivial difference compared with the EA with no alignment, while for videos under negative emotion (first column,  Figure~\ref{fig:delvin_misalignments_scatterplot}), the difference is much bigger. This suggests a stronger time-warping effect for negative emotions, which is consistent with psychological research indicating better recognition of positive emotions \citep{Bandyopadhyay2021}. Figure \ref{fig:Delvin_pairwise_CI}) shows that the penalized SRVF methods lead to estimated EA with a significantly smaller mean than those obtained from unpenalized SRVF in all four videos. In three out of four videos, the mean EA obtained from the penalized SRVF is also significantly larger than those from no alignment, and generally differs significantly from those obtained under the optimal fixed delay method. In addition, Figure S1 in the Supplementary Materials shows the correlations among the EA measures obtained by different alignment methods. Although they are positively correlated with others, they are not equivalent and our proposed alignment method can lead to improved inference in a downstream analysis.

We also examined the associations between perceiver-specific trait positive emotion and their EA. Trait positive emotion reflects a perceiver's stable tendency to experience positive emotions across diverse situations and over time, and it is typically associated with greater sociability, prosocial behavior, and openness \citep{devlin2014not}. To make this analysis consistent with the approach adopted by \citet{devlin2014not}, we fitted a simple linear regression model with the Fisher transformed EA measure as the outcome and trait positive emotion as the predictor. Table \ref{tab:EA-variable} summarizes the estimated slope of each regression. It reveals that the penalized SRVF method consistently yields larger absolute coefficient estimates than the no-alignment approach across all four videos. This pattern is not consistently observed with the unpenalized SRVF or the fixed delay methods. %\textcolor{blue}{Furthermore, when no alignment is conducted, a high EA is negatively associated with a high trait positive affect in the high-intensity negative video, but positively associated in the low-intensity positive video. Jing: what is the point of this sentence?} Furthermore, all alignment methods detect a significant negative association in the high-intensity positive video, whereas the no-alignment method does not. 
These findings suggest that failing to properly address misalignment may obscure important relationships between EA and perceiver characteristics.
%We further investigated whether the amount of warping $d_p$ is associated with the trait positive emotion covariate. Note that this analysis is not possible when no alignment is conducted because it assumes zero warping.  All the alignment methods lead to non-significant association with the trait positive affect; this may suggest temporal warping is another distinctive feature of perceivers' in the perception process \textcolor{blue}{Jing: why so? There is one significant result}.
\begin{table}[t]
\centering
\vspace{-2mm}
\caption{Estimated coefficients for Trait Positive Affect as a predictor of Empathic Accuracy as measured by different alignment methods. Standard errors are included in parentheses and * indicates significance at the significance level $5\%$.}
\begin{tabular}{l llll}
\toprule
  & High Negative & Low Negative & High Positive & Low Positive\\
\hline
\addlinespace
%\multicolumn{5}{c}{\it EA measure } \\[.5em]
No Alignment & -0.013 (0.005)* & -0.003 (0.002) & -0.011 (0.006) & 0.012 (0.005)*\\
Fixed Delay & -0.017 (0.006)* & -0.004 (0.003) & -0.017 (0.007)* & 0.005 (0.004)\\
Unpenalized SRVF & -0.009 (0.005)* & -0.002 (0.003) & -0.024 (0.011)* & 0.019 (0.009)*\\
Penalized SRVF (8s) & -0.014 (0.005)* & -0.004 (0.003) & -0.016 (0.008)* & 0.013 (0.005)*\\
%\multicolumn{5}{c}{\it Temporal Warping} \\[.5em]
%No Alignment & - & - & - & -\\
%Fixed Delay & 0.001 (0.001) & -0.001 (0.001) & 0.002 (0.002) & -0.001 (0.002)\\
%Unpenalized SRVF & 0.000 (0.001) & 0.002 (0.001) & 0.002 (0.001) & -0.002 (0.001)\\
%Penalized SRVF & 0.000 (0.000) & 0.000 (0.001) & 0.001 (0.001) & 0.000 (0.001)\\
\bottomrule
\label{tab:EA-variable}
\end{tabular}
\end{table}

\subsection{Study on Music Empathy}
\label{sec: data_music}
%\subsection{Data and model}
%\citet{tabak2022initial} conducted an EA study of music, where their stimuli consisted of nine music recordings (1–2 minutes each) that were composed to express the three primary emotions of joy/happiness, sadness, and anger. Each primary emotion corresponds to $J = 3$ recordings. 
\citet{tabak2022initial} conducted an EA study investigating three primary emotions: joy/happiness, sadness, and anger ($I = 3$). For each emotion, three original solo piano pieces ($J = 3$) were composed and performed by experienced musicians. These pieces were designed to evoke the target emotions within familiar musical styles (classical, popular, and jazz). Both musicians (as targets) and 123 participants (as perceivers) rated the emotional content of each piece on a 9-point scale. As with the previous dataset, we preprocessed the data by smoothing and interpolating rating functions.

Unlike the correlation-based approach, \citet{tabak2022initial} employed a linear mixed-effect model for a more nuanced analysis of EA. This model decomposed perceiver responses into three latent factors: bias, discrimination, and variance. Bias represented the systematic deviation between perceiver and target ratings, while discrimination captured a perceiver's sensitivity to changes in the target's expressed emotion. Finally, variance accounted for random noise in perceiver ratings. 

Within each group of emotion ($i=1,\ldots, I$),  let $x_j(\cdot)$ and $y_j(\cdot)$ be the target and a perceiver's ratings, respectively, for the $j$th stimulus. \citet{tabak2022initial} proposed the following linear mixed-effect model to describe the relation between $x_j(\cdot)$ and $y_j(\cdot)$:
\begin{equation}
y_{jk} = \beta_0 + \beta_1 x_{jk} + b_{0j} + b_{1j} x_{jk} + \varepsilon_{jk}, ~j=1,\ldots,J,~ k=1, \ldots, T_j,
\label{eq:LMM1}
\end{equation}
where $y_{jk} = y_j(t_k)$ and $x_{jk} = x_j(t_k)$ are the perceiver and target's respective ratings at the $k$th time point, and $T_j$ is the number of points for the $j$th stimuli. The (fixed) intercept $\beta_0$ and slope $\beta_1$ represent a perceiver's mean bias and discrimination ability across all the $J$ stimuli, respectively. The random intercept $b_{0j}$, random slope $b_{1j}$, and the random noise $\varepsilon_{jk}$ are assumed to follow a normal distribution with zero mean and respective variance component $\sigma_{b_0}^2, \sigma_{b_1}^2$ and $\sigma^2$, which represents the variability of bias, discrimination, and random noise across different stimuli. This model treats ratings as discrete points and does not account for potential misalignments between perceiver and target responses.

To address this limitation, we integrated an alignment step into the model framework.  Treating the observed ratings as sampled points from corresponding functions, we applied and compared penalized and unpenalized time-warping SRVF alignments to account for potential misalignments. Let $\tilde{y}_j(t) =  y_j \circ \hat{\gamma}_j(t)$ be the estimated aligned function with $\hat{\gamma}_j(t)$ being an estimated \revision{inverse} warping function from aligning $y_j(\cdot)$ with $x_j(\cdot)$, we then modeled
\begin{equation}
\tilde{y}_j(t) = \beta_0 + \beta_1 x_j(t) + b_{0j} + b_{1j}x_j(t) + \varepsilon_j(t), ~b_{0j} \sim N(0, \sigma_{b_0}^2), ~ b_{1j} \sim N(0, \sigma_{b_1}^2),
\label{eq: LMM}
\end{equation}
where $\beta_0, \beta_1$, $\sigma_{b_0}^2$, $\sigma_{b_1}^2$ and $\sigma^2$ in Model \eqref{eq: LMM} maintain the same interpretations as in Model \eqref{eq:LMM1}.
We fitted Model \eqref{eq: LMM}  for each perceiver and primary emotion. Using the \texttt{lme4} package in R \citep{douglas2015lme4}, we employed restricted maximum likelihood estimation to obtain parameter estimates $\hat\Psi = (\hat\alpha, \hat\beta, \hat\sigma_{b_0}^2, \hat\sigma_{b_1}^2, \hat\sigma^2)$ and best linear unbiased predictions (BLUPs) of random effects $\hat{b}_{0j}$ and $\hat{b}_{1j}$ for $j=1,\ldots, J$. To assess the impact of time warping, we compared parameter estimates  $\hat\Psi$ under no alignment (i.e., $\gamma_{id}$), the unpenalized SRVF ($\hat{\gamma}_{u}$), and the penalized SRVF alignment ($\hat{\gamma}_p)$, setting the penalty threshold at $\nu = 8$ seconds for the penalized alignment. Results for 6- and 10-second thresholds are provided in Section~S3.2 of the Supplementary Material.

To assess model fit, we computed two metrics: average warping and average goodness of fit across all $J$ tasks. The first metric, average Fisher-Rao distance, quantifies the mean warping magnitude relative to the identity warping: $\Bar{d}_p = J^{-1} \sum_{j=1}^{J} \cos^{-1}\left( \int_{0}^{1}\sqrt{{\hat{\gamma}}^\prime_j(t)} dt \right)$. A higher $\Bar{d}_p$ indicates greater warping. The second metric measures the vertical distance between the estimated aligned response function and the fitted value function. Specifically, {letting $\hat{y}_j(\cdot) = \hat\beta_0 + \hat\beta_1 x_j(\cdot) + \hat{b}_{0j} + \hat{b}_{1j} x_j (\cdot)$, this vertical distance is calculated to be the $\mathbb{L}^2$ distance between the estimated aligned response $\tilde{y}_j$ and the fitted value function $\hat{y}_j$, i.e.,  
$\sum_{j=1}^{J} \Vert \tilde{y}_j - \hat{y}_j \Vert_2^2
$,} where a lower value signifies a better model fit. %We note that we used $\mathbb{L}^2$ distance between functions instead of SRVF distances because the initial values of functions are different which cannot be captured by the SRVF distances.

%To compare the model fit, we computed two metrics, one representing the average amount of warping, and the other repre$J$senting the average goodness of fit across all  tasks. For the first metric, we computed the average Fisher-Rao metric between the estimated warping functions and the identity warping, $\Bar{d}_p = J^{-1} \sum_{j=1}^{J} \cos^{-1}\left( \int_{0}^{1}\sqrt{{\hat{\gamma}}^\prime_j(t)} dt \right)$, which is similar to the metric as defined in Section \ref{subsection:pen}. A higher value for this metric implies a higher amount of warping. For the second metric, we computed the vertical distance between the aligned response function and the fitted value function. Specifically, let $\hat{y}_j(\cdot) = \hat\beta_0 + \hat\beta_1 x_j(\cdot) + \hat{b}_{0j} + \hat{b}_{1j} x_j (\cdot)$, this vertical distance is calculated as$\sum_{j=1}^{J} \Vert \tilde{y}_j - \hat{y}_j \Vert_2^2$. A lower vertical distance implies a better model fit. 
\begin{figure}[t]
    \centering
    \includegraphics[width=\linewidth, page = 1]{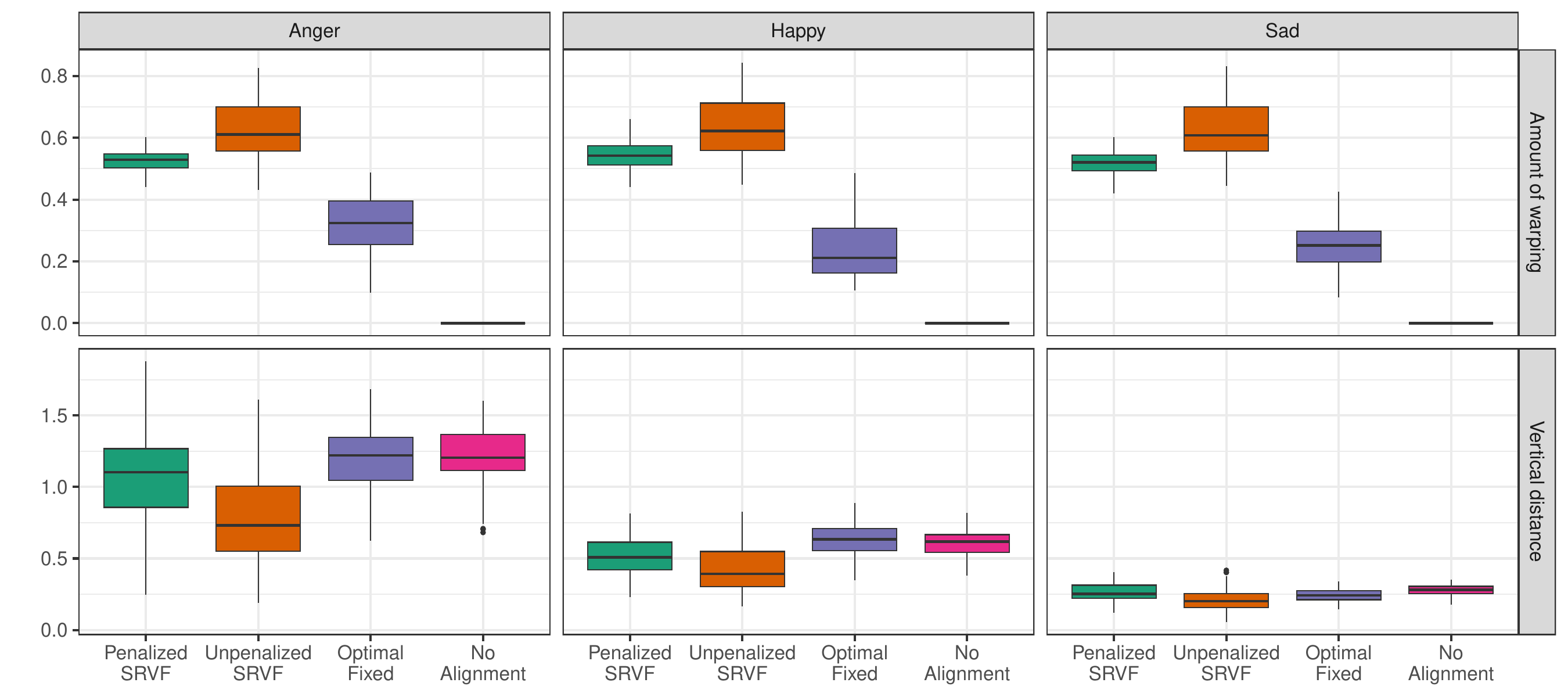}
    \caption{Boxplots of the metrics for the average amount of warping (top row) and goodness of fit (bottom row) of the estimated models for the three sets of music recordings. The penalized SRVF alignment was conducted using the threshold $\nu = 8s$.    }
    \label{fig:phasesvsRSS}
\end{figure}

Figure \ref{fig:phasesvsRSS} reveals that the no alignment model exhibits significantly inferior fit compared to the other two approaches (all the $p$-values for comparing pairwise mean differences are close to zero, see Table S4 in the Supplementary Materials). This underscores the importance of addressing misalignment between perceiver and target ratings to prevent model underfitting. Similar to our simulation findings, the unpenalized SRVF alignment demonstrates overfitting, sacrificing model fit for excessive warping. \revision{In contrast, the penalized alignment method provides a reasonable compromise between these extremes, enhancing model fit while mitigating overfitting through judicious penalty application.}

\begin{figure}[p]
    \begin{subfigure}{\textwidth}
    \centering
    \includegraphics[width=.8\linewidth, page = 1]{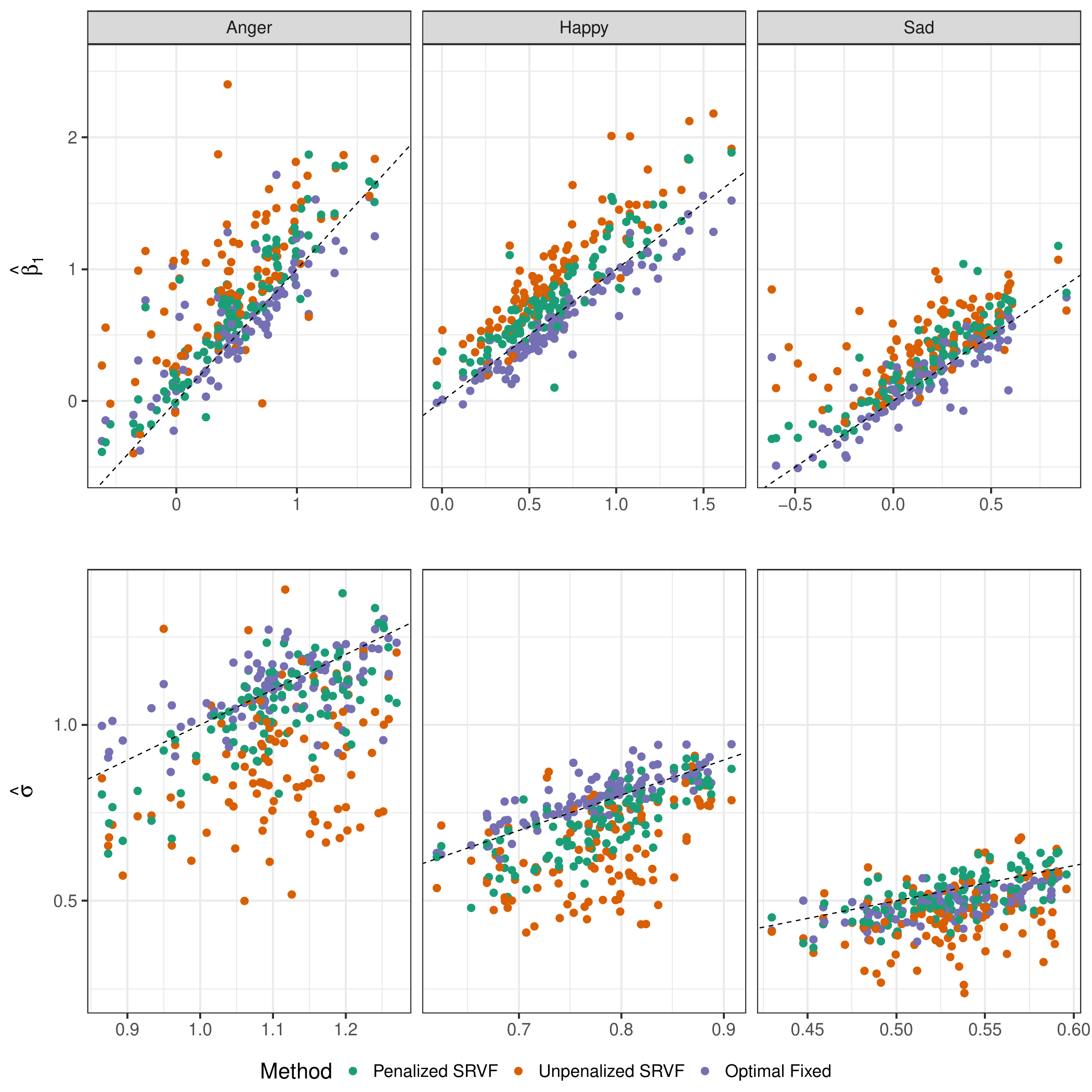}  
     \caption{Scatterplots of estimates for the fixed effect discrimination $\beta_1$ (top) and random noise standard deviation $\sigma$ (bottom). In each plot, the horizontal axis represents the estimate when no alignment is conducted, and the vertical axis represents the estimates under unpenalized SRVF (orange) and penalized SRVF with $\nu = 8s$ (green). The dashed line represents the 45-degree line.   }
         \label{fig:parameterest}
    \end{subfigure}
    \begin{subfigure}{\textwidth}
        \centering
    \includegraphics[width=.9\linewidth]{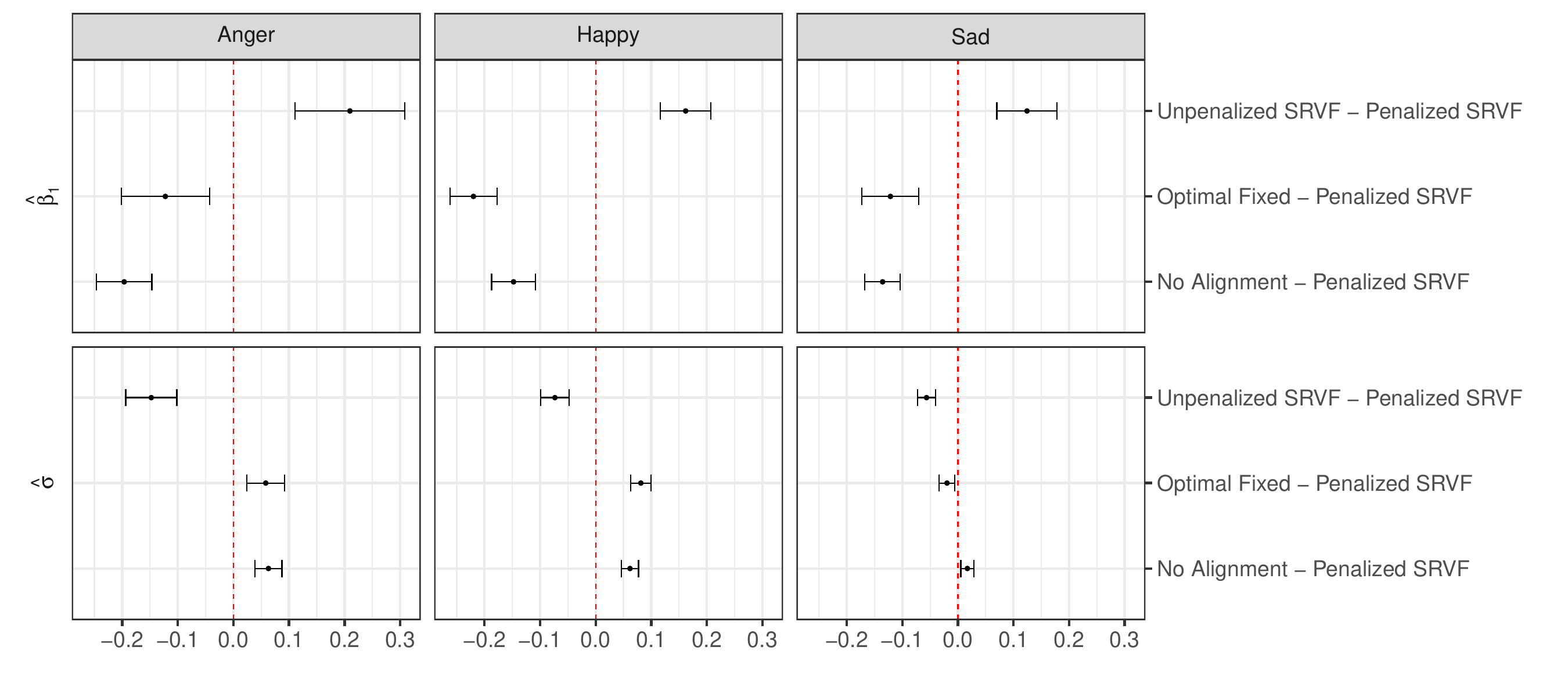}
    \caption{Pairwise confidence intervals for the mean difference between the estimates obtained from unpenalized SRVF, optimal fixed, and no alignment and the proposed penalized SRVF method. }
    \label{fig:tabak_pairwise_CI}
\end{subfigure}
   \caption{Results for parameter estimates in the music EA study}

\end{figure}

Figure \ref{fig:parameterest} compares parameter estimates (aligned vs. unaligned) for the fixed effect discrimination ($\hat{\beta}_1$) and random noise standard deviation ($\hat{\sigma}$) across the three emotion groups, and Figure \ref{fig:tabak_pairwise_CI} shows the pairwise confidence intervals in the mean estimates of the penalized SRVF method against other alignment methods. In the top row of both figures, while both unpenalized SRVF and penalized SRVF alignment methods tend to increase the discrimination estimates $\hat\beta_1$, the optimal fixed delay tends to decrease it compared to the no alignment. %Hence, the optimal fixed delay appears to contradict with what we have expected: When no alignment is conducted, the estimated discrimination effect $\hat\beta_1$ can be viewed as the true $\beta_1$ contaminated by measurement error. 
However, similar to the social empathy study in section \ref{sec:data_social}, the unpenalized SRVF alignment increases this discrimination estimate much more than the penalized SRVF, making the unpenalized SRVF more prone to overfit. This conclusion is further evidenced in the bottom row, where the estimated standard deviation $\hat\sigma$ from the unpenalized SRVF method is substantially smaller than that from the other two alignment methods. %Except for the Sad group, the optimal fixed delay tends to increase the estimated variability compared to no alignment estimates, which contradicts what we have expected. \textcolor{blue}{Jing: why this contradicts what we have expected? Is it OK we remove the last sentence?} %  \textcolor{blue}{Overall, the performance of the optimal fixed delay alignment method in this music EA study may be aligned with what we have expected .... The proposed SRVF method is the most reliable among the four methods,... (Jing: what do you mean by this?)}

%Regarding the fixed effect discrimination ($\hat{\beta}_1$), unaligned responses can be viewed in a certain way as aligned responses contaminated by measurement error. 

%he attenuation bias in $\hat{\beta}_1$ estimates is reduced, leading to increased values as depicted in Figure s first row. However, the excessive warping associated with unpenalized alignment amplifies this effect.
 %Both alignment methods reduce $\hat{\sigma}$ compared to the no-alignment condition, as misaligned rating sequences inflate random noise. Aligned responses exhibit stronger associations with target responses, decreasing unexplained variability. While both alignments reduce $\hat{\sigma}$, the unpenalized alignment achieves a more substantial reduction due to over-alignment.

\section{Discussion}
\label{sec:discussion}

In emotional perception research, misalignment caused by complex cognitive decoding processes and the time needed to enact a behavioral response is a well-established phenomenon. In EA studies, the discrepancy between the perceiver's observed rating and the target's rating is influenced by both the misalignment not due to lack of EA and the psychologically meaningful disagreement resulting from lack of EA. Yet, most of the conventional EA studies either ignore this kind of misalignment or apply an oversimplified fixed delay for adjustment, where both options can lead to biased results. This study introduces a novel, flexible
approach using a new \revision{constrained optimization problem based on the SRVF representation of the functions}  to reduce the misalignment, which varies from individual to individual. \revision{Considering realistic conditions of the warping process, our simulation studies demonstrate that the proposed penalized SRVF alignment method provides improved estimates of the true EA measure compared to existing approaches. In two case studies on social and music empathy, this method yields plausible EA measures, which subsequently reveals more potential associations between EA and perceivers’ characteristics. }

%Misalignment is a well-established phenomenon in emotional perception research, yet there remains a dearth of rigorous statistical methods to correct it. Neglecting misalignment in EA data analysis can lead to significantly biased results. Conventional approaches, such as accounting for fixed delays in responses, oversimplify the complex nature of misalignment. This study introduces a novel, flexible approach using penalized warping functions to address the multifaceted nature of misalignment in emotion perception studies.

%While warping functions have been widely employed to correct misalignment in fields like physics and biology, where objective benchmarks exist, their application to the abstract and subjective domain of human perception is, to the best of our knowledge, largely unexplored. The challenges inherent in accurately measuring perceptual constructs have often led to the neglect of misalignment in cognitive research. This study represents a significant step toward addressing this critical issue.

The proposed penalized alignment approach offers several advantages. 1) Individualized adjustments: It tailors alignment to unique patterns of misalignment for each perceiver.
2) Prevention of over-alignment: It incorporates a natural constraint on the extent of allowable warping. 3) Simplicity and interpretability: The penalty term can be easily set up by using the maximum delay in perceivers' responses, where it is straightforward to use empirical evidence and expert opinion. 
\revision{Moreover, EA studies often vary in context and focus. In situations where reaction time is less critical, such as listening to a friend’s long story, a broader penalization window may be appropriate. Conversely, in high-stakes scenarios like heated arguments, a narrower window can better reflect the urgency of responses. This flexibility enables researchers to tailor penalization parameters to the specific demands of each study. By integrating these features, our approach enhances the accuracy of downstream EA analyses, including correlational studies and complex linear mixed models.}
%By providing a flexible and robust method for correcting misalignment not due to EA, our approach empowers researchers to conduct more accurate downstream EA analyses, such as correlational studies and complex linear mixed models. 
%We have made our code publicly available (XXX) to facilitate a widespread adoption of the proposed method

The core component in our proposed method, the warping functions, has been widely employed to correct misalignment in fields like physics and biology where objective benchmarks exist. To the best of our knowledge, the application of warping functions to the abstract and subjective domain of human perception is unexplored. In this study, we have demonstrated their effectiveness and flexibility in adjusting individually varying misalignment across different types of emotional stimuli (visual and audio). It further expands the potential for using warping functions in new research areas. 

Future research could focus on several key areas. One potential direction is to model the similarity of
warping functions of the same individual across different stimuli by introducing random effects. Another area of interest could be to develop a new EA alignment method by incorporating additional data, such as the functional magnetic resonance imaging (fMRI) blood-oxygen-level-dependent (BOLD) signals
of targets and perceivers during rating assessments, which could help detect true emotional changes. \revision{Although our method accurately identified simulated noise and showed favorable psychometric characteristics, we caution against interpreting the corrected scores from our method as definitive indicators of empathic accuracy devoid of all measurement error. Future work is needed to explicitly test the extent to which the penalized alignment approach can distinguish measurement noise from meaningful differences in EA, for example, by experimentally manipulating whether participants can pause the video to make ratings or by varying the cognitive load placed on participants.}
Finally, we note that these analyses
were exploratory in nature. Given the methodological flexibility of the proposed method and
the number of analytic decisions involved (e.g., the upper limit of warping functions), future
work can aim to replicate these findings using pre-registered designs to increase confidence in the robustness of the results provided by the penalized SRVF alignment method.

\paragraph{Supplementary Material}
The supplementary materials contain a proof of Lemma 3.1, additional results for the simulations, and the two data applications. 

\paragraph{Data Availability Statement}
The data used in the two case studies, along with the R codes that implement the penalized SRVF alignment method and produce numerical results in the paper, \revision{are available at the GitHub repository: \url{https://tinyurl.com/5dba3fzh}}.

\bibliographystyle{apalike}
\bibliography{references}

\appendix

\newpage
\noindent \textbf{\Large SUPPLEMENTARY MATERIALS}
\section{Proof of Lemma 3.1}

Let $\Gamma_I^*$ be the set of warping functions under penalized warping of (3.2) in the paper, $\Gamma_I^*=\{\gamma:[0,1]\rightarrow[0,1]$, $\gamma(0)=0$, $\gamma(1)=1$, $\gamma$ is invertible, $\gamma$ and $\gamma^{-1}$ are smooth, $\sup|\gamma-\gamma_{id}|\leq \nu \}$.
By representing $\gamma \in \Gamma_I^*$ as its SRVF $\sqrt{{\gamma}^\prime}$, the space $\Gamma_I^*$ maps to $\mathcal{Q}^*=\{\sqrt{{\gamma}^\prime}:[0,1] \mapsto \mathbb{R}_{\geq 0} \mid \int_0^1 \left(\sqrt{{\gamma}^\prime(s)}\right)^2 dt = 1,\left( \int_0^t \left(\sqrt{{\gamma}^\prime(s)}\right)^2ds-t \right)^2 \leq \nu^2 \text{ for } 0\leq t \leq 1 \}$.
Here, $\Gamma_I^* \subset \Gamma_I$ and $\mathcal{Q}^* \subset \mathcal{Q}=\{\sqrt{\Dot{\gamma}}:[0,1] \mapsto \mathbb{R}_{\geq 0} \mid \int_0^1 \left(\sqrt{\Dot{\gamma}(s)}\right)^2 dt = 1 \}$. 
Because $\mathcal{Q}$ is a subset of the positive orthant of the unit Hilbert sphere $\mathbb{S}^+_{\infty}$, $\mathcal{Q}^*$ also is a subset of $\mathbb{S}^+_{\infty}$. Therefore, the SRVF of warping functions from the penalized SRVF are the elements of a unit Hilbert sphere, and their distance $d_p$ is the arc-length distance. It can be approximated by $d_p(x,y) \approx \cos^{-1}\left(\int_0^1 \sqrt{{\hat{\gamma}^\prime}_{p}(t)} dt\right)$.

\section{Supplements to Simulation Study}

\begin{table}[!ht]
\caption{
    Performance of different alignment methods in the simulation studies under different warping limits $\eta$, measured by the $\mathbb{L}^2$ distance $d_a$ between the aligned perceiver $\hat{y}$ and the true latent perceiver $a$, and the $(10\times)$ bias of the estimated correlation between the true latent  perceiver and the target.  The lowest absolute bias and the lowest $d_a$ are highlighted for each row. Standard errors are included in the parentheses.  }
    \label{tab:sim_summary_parameters}
\begin{tabular}[t]{lllcccccc}
\toprule
$\eta$ & Video & Metric & Pen. SRVF & $\mathbb{L}^2$ SRVF & Unpen. SRVF & Opt. Fixed & No Alignment\\
\midrule
6 &  \multirow[t]{2}{1cm}{High Neg} & $d_a$ & \textbf{3.74 (1.02)} & 6.63 (3.58) & 11.72 (5.13) & 4.64 (1.81) & 4.47 (1.27)\\
 &  & Bias & \textbf{0.03 (0.28)} & 0.67 (1.00) & 1.22 (1.41) & -0.10 (0.51) & -0.09 (0.31)\\
\addlinespace
 & \multirow[t]{2}{1cm}{Low Neg} & $d_a$ & {6.95 (4.38)} & \bfseries 3.80 (2.35) & 9.01 (4.13) & 8.00 (5.22) & 8.96 (5.36)\\
 &  & Bias & {-0.63 (1.43)} & \bfseries 0.61 (1.10) & 1.35 (1.70) & -0.80 (2.09) & -1.30 (1.62)\\
\addlinespace
 & \multirow[t]{2}{1cm}{High Pos} & $d_a$ & \textbf{4.47 (1.20)} & 6.65 (3.87) & 11.83 (5.53) & 6.112 (2.94) & 4.69 (1.27)\\
 &  & Bias & \textbf{-0.00 (0.59)} & 0.67 (1.24) & 0.80 (2.27) & -0.08 (1.22) & -0.08 (0.61)\\
\addlinespace
 & \multirow[t]{2}{1cm}{High Neg} & $d_a$ & \textbf{4.48 (1.35)} & 4.52 (3.03) & 7.94 (3.48) & 6.00 (3.20) & 5.07 (1.40)\\
 &  & Bias & \textbf{-0.06 (0.47)} & 0.60 (0.85) & 0.74 (1.20) & -0.41 (1.21) & -0.20 (0.49)\\
\addlinespace
$\Gamma(6,1)$ & \multirow[t]{2}{1cm}{High Neg} & $d_a$ & \textbf{3.88 (1.71)} & 6.37 (3.45) & 11.25 (4.84) & 4.72 (2.41) & 4.49 (1.85)\\
 &  & Bias & \textbf{0.04 (0.31)} & 0.60 (0.91) & 1.18 (1.35) & -0.16 (0.56) & -0.08 (0.33)\\
\addlinespace
 & \multirow[t]{2}{1cm}{Low Neg} & $d_a$ & {6.70 (4.51)} & \bfseries 4.05 (2.68) & 9.27 (4.09) & 7.79 (4.84) & 8.74 (5.44)\\
 &  & Bias & \textbf{-0.52 (1.46)} & 0.66 (1.22) & 1.43 (1.71) & -0.63 (1.95) & -1.14 (1.66)\\
\addlinespace
 & \multirow[t]{2}{1cm}{High Pos} & $d_a$ & \textbf{4.50 (1.54)} & 6.67 (3.87) & 12.38 (5.97) & 6.21 (2.98) & 4.70 (1.63)\\
 &  & Bias & \textbf{-0.04 (0.59)} & 0.59 (1.22) & 0.61 (2.43) & -0.14 (1.24) & -0.12 (0.60)\\
\addlinespace
 & \multirow[t]{2}{1cm}{High Neg} & $d_a$ & {4.54 (1.70)} & \bfseries 4.41 (2.76) & 8.33 (3.37) & 5.81 (3.37) & 5.16 (1.73)\\
 &  & Bias & \textbf{-0.07 (0.50)} & 0.59 (0.90) & 0.71 (1.29) & -0.27 (1.20) & -0.22 (0.52)\\
 \addlinespace[1em]

10& \multirow[t]{2}{1cm}{High Neg} & $d_a$ & \textbf{5.94 (1.78)} & 6.63 (3.58) & 11.58 (5.30) & 8.36 (4.21) & 6.88 (1.88)\\
 &  & Bias & \textbf{0.01 (0.53)} & 0.67 (1.00) & 1.23 (1.43) & -0.52 (1.27) & -0.23 (0.52)\\
\addlinespace
 & \multirow[t]{2}{1cm}{Low Neg} & $d_a$ & {9.80 (5.83)} & \bfseries 3.80 (2.35) & 9.20 (4.06) & 10.67 (5.54) & 12.47 (6.72)\\
 &  & Bias & {-0.87 (2.07)} & \bfseries 0.61 (1.10) & 1.43 (1.83) & -1.14 (2.57) & -1.98 (2.35)\\
\addlinespace
 & \multirow[t]{2}{1cm}{High Pos} & $d_a$ & {6.92 (1.94)} & \bfseries 6.65 (3.87) & 11.99 (5.17) & 9.31 (3.88) & 7.25 (2.03)\\
 &  & Bias & \textbf{-0.07 (1.00)} & 0.67 (1.24) & 0.70 (2.56) & -0.49 (2.15) & -0.22 (1.02)\\
\addlinespace
 & \multirow[t]{2}{1cm}{High Neg} & $d_a$ & {7.12 (2.32)} & \bfseries 4.52 (3.03) & 8.17 (3.24) & 9.48 (4.77) & 7.73 (2.18)\\
 &  & Bias & \textbf{-0.18 (0.80)} & 0.60 (0.85) & 0.69 (1.12) & -0.90 (2.31) & -0.39 (0.81)\\
\addlinespace
$\Gamma(10,1)$  &\multirow[t]{2}{1cm}{High Neg} & $d_a$ & \textbf{5.69 (2.12)} & 6.37 (3.45) & 11.25 (4.69) & 7.78 (3.95) & 6.62 (2.25)\\
 &  & Bias & \textbf{0.04 (0.50)} & 0.60 (0.91) & 1.20 (1.34) & -0.42 (1.13) & -0.19 (0.51)\\
\addlinespace
 & \multirow[t]{2}{1cm}{Low Neg} & $d_a$ & \textbf{9.27 (5.51)} & 4.05 (2.68) & 9.07 (4.08) & 10.57 (5.84) & 11.97 (6.54)\\
 &  & Bias & \textbf{-0.77 (2.01)} & 0.66 (1.22) & 1.34 (1.61) & -1.21 (2.67) & -1.88 (2.33)\\
\addlinespace
 & \multirow[t]{2}{1cm}{High Pos} & $d_a$ & {6.76 (2.17)} & \bfseries 6.67 (3.87) & 11.68 (5.72) & 8.97 (3.75) & 7.10 (2.27)\\
 &  & Bias & \textbf{-0.05 (1.00)} & 0.59 (1.22) & 0.79 (2.25) & -0.45 (2.05) & -0.21 (1.02)\\
\addlinespace
 & \multirow[t]{2}{1cm}{High Neg} & $d_a$ & {7.00 (2.48)} &\bfseries 4.41 (2.76) & 8.34 (3.61) & 9.16 (4.68) & 7.66 (2.43)\\
 &  & Bias & \textbf{-0.10 (0.88)} & 0.59 (0.90) & 0.63 (1.22) & -0.94 (2.28) & -0.32 (0.89)\\
 \bottomrule
\end{tabular}
\end{table}
\FloatBarrier

\section{Supplementary Results for Data Applications}

\subsection{Data application: Social empathy}

This section contains additional results for the first data application, including the estimates of the amount of warping and correlation between the aligned perceivers and the targets under different thresholds.   

\tabcolsep .4em
\begin{table}[H]
\caption{Mean (standard deviation) of the estimated amount of warping and correlation between aligned perceiver and the target across $122$ perceivers in the social empathy study, under different alignment methods, including no alignment, unpenalized SRVF, and penalized SRVF with thresholds $\nu \in \{6, 8, 10\}$ seconds.}
\begin{tabular}{llllccc}
\toprule
Video & Metric & \multirow{2}{2cm}{\centering No alignment }& \multirow{2}{1.5cm}{\centering Unpen.  SRVF} & \multicolumn{3}{c}{Pen. SRVF} \\
\cmidrule{5-7}
& & & & $\nu=6$ & $\nu=8$ & $\nu=10$ \\
\midrule
\multirow[t]{2}{1.2cm}{High Neg} & Correlation & 0.51 (0.14) & 0.68 (0.11) & 0.58 (0.15) & 0.59 (0.16) & 0.58 (0.17)\\
 & Amount of Warping & 0.00 (0.00) & 0.66 (0.09) & 0.50 (0.05) & 0.55 (0.05) & 0.57 (0.05)\\
\addlinespace

\multirow[t]{2}{1.2cm}{High Pos}  & Correlation & 0.73 (0.20) & 0.85 (0.13) & 0.75 (0.19) & 0.75 (0.19) & 0.75 (0.19)\\

& Amount of Warping & 0.00 (0.00) & 0.66 (0.10) & 0.50 (0.03) & 0.52 (0.04) & 0.53 (0.04)\\

\addlinespace

\multirow[t]{2}{1.2cm}{Low Neg} & Correlation & 0.89 (0.15) & 0.85 (0.26) & 0.91 (0.15) & 0.91 (0.15) & 0.91 (0.15)\\

 & Amount of Warping & 0.00 (0.00) & 0.65 (0.08) & 0.49 (0.05) & 0.52 (0.05) & 0.53 (0.05)\\

\addlinespace

\multirow[t]{2}{1.2cm}{Low Pos} & Correlation & 0.46 (0.28) & 0.68 (0.44) & 0.50 (0.29) & 0.52 (0.30) & 0.54 (0.31)\\

 & Amount of Warping & 0.00 (0.00) & 0.64 (0.07) & 0.49 (0.04) & 0.53 (0.05) & 0.54 (0.05)\\
\bottomrule
\end{tabular}            
\end{table}

\begin{table}
\caption{Point estimates, 95\% confidence intervals and $p$-values for the difference in the estimated amount of warping obtained by the proposed penalized SRVF with warping limit $\nu=8s$ versus that obtained by the other alignment method.}
\centering
\begin{tabular}[t]{llrrrr}
\toprule
Video & Pair & Est & Low CI & Upp CI & $p$-value\\
\midrule
\multirow[t]{3}{1.5cm}{High Neg} & No Alignment - Penalized SRVF & -0.52 & -0.53 & -0.51 & 0.00\\
& Optimal Fixed - Penalized SRVF & -0.33 & -0.35 & -0.31 & 0.00\\
& Unpenalized SRVF - Penalized SRVF & 0.14 & 0.12 & 0.16 & 0.00\\
\addlinespace
\multirow[t]{3}{1.2cm}{High Pos} & No Alignment - Penalized SRVF & -0.55 & -0.56 & -0.53 & 0.00\\
& Optimal Fixed - Penalized SRVF & -0.33 & -0.36 & -0.31 & 0.00\\
& Unpenalized SRVF - Penalized SRVF & 0.11 & 0.08 & 0.13 & 0.00\\
\addlinespace
\multirow[t]{3}{1.2cm}{Low Neg} & No Alignment - Penalized SRVF & -0.52 & -0.53 & -0.50 & 0.00\\
 & Optimal Fixed - Penalized SRVF & -0.26 & -0.30 & -0.23 & 0.00\\
& Unpenalized SRVF - Penalized SRVF & 0.13 & 0.11 & 0.15 & 0.00\\
\addlinespace
\multirow[t]{3}{1.2cm}{Low Pos} & No Alignment - Penalized SRVF & -0.53 & -0.54 & -0.52 & 0.00\\
& Optimal Fixed - Penalized SRVF & -0.26 & -0.29 & -0.22 & 0.00\\
& Unpenalized SRVF - Penalized SRVF & 0.10 & 0.08 & 0.13 & 0.00\\
\bottomrule
\end{tabular}
\end{table}

\begin{figure}[ht]
 \centering
    \includegraphics[width=\linewidth]{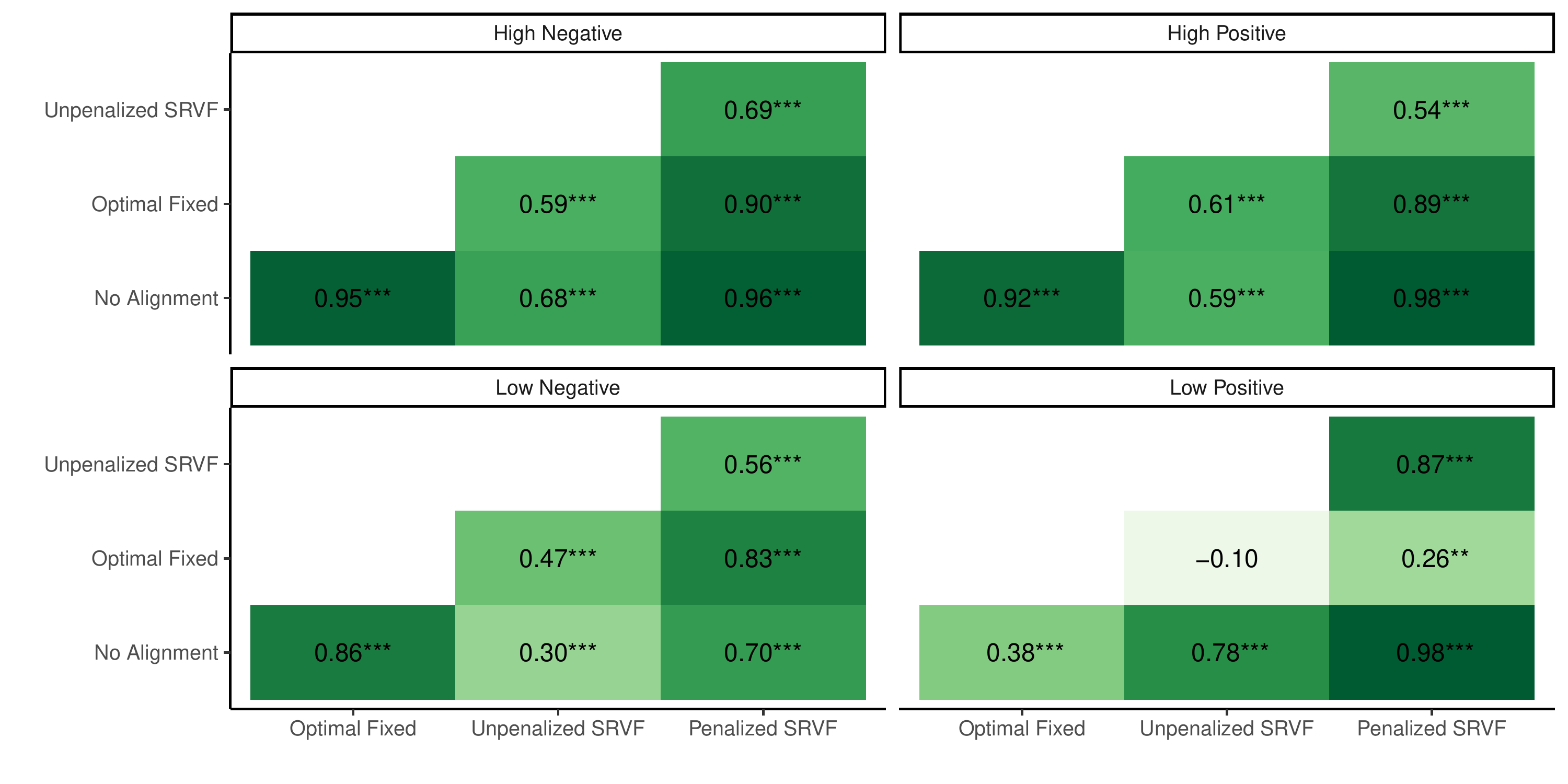}
        \caption{Correlations among EA measures obtained by different alignment methods in the social EA study}
\end{figure}

\begin{figure}[H]
    \centering
    \includegraphics[width=\linewidth, page=3]{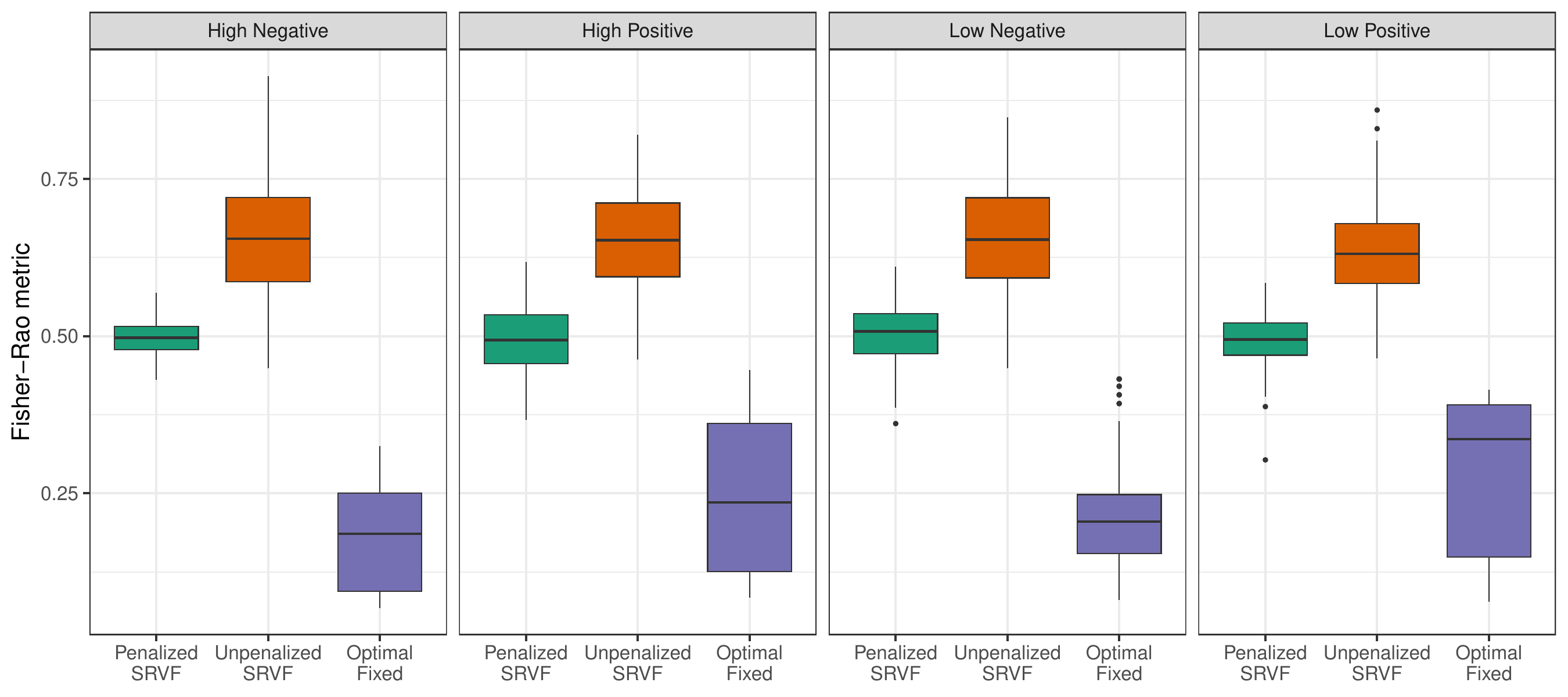}
  \caption{Boxplots for the estimated amount of warping, as measured by the Fisher-Rao metric between the identity warping $\gamma_{id}$ and the estimated warping function using unpenalized SRVF and penalized SRVF method with $\nu = 6s$ for each video. }
    
\end{figure}

\begin{figure}[H]
    \centering
        \includegraphics[width=\linewidth, page=1]{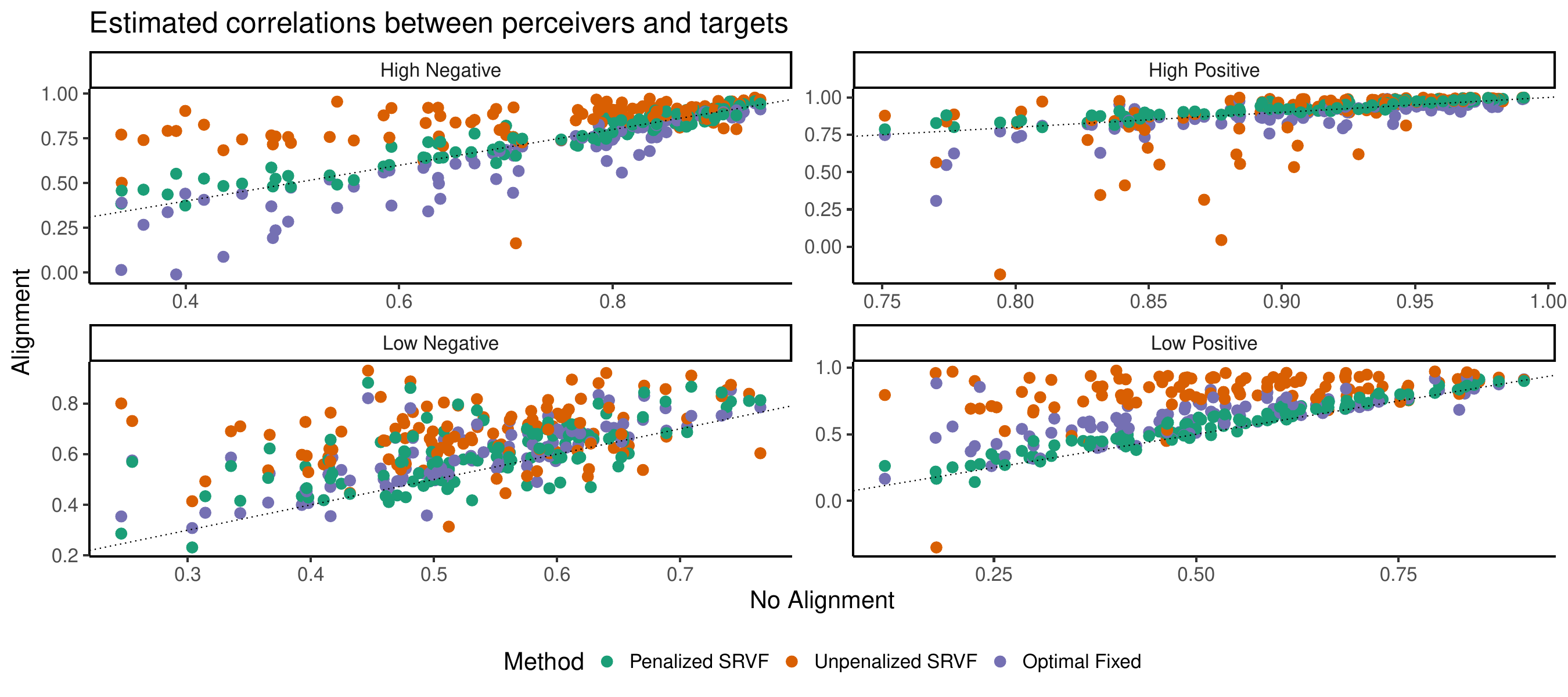}
    \caption{Scatterplots of correlation between target's ratings and each perceiver's ratings for each video. In each plot, the horizontal axis represents the correlation when no alignment is conducted, and the vertical axis represents the estimates under unpenalized SRVF (orange) and penalized SRVF with $\nu = 6s$ (green). }

\end{figure}

\begin{figure}[H]
    \centering
    \includegraphics[width=\linewidth, page=6]{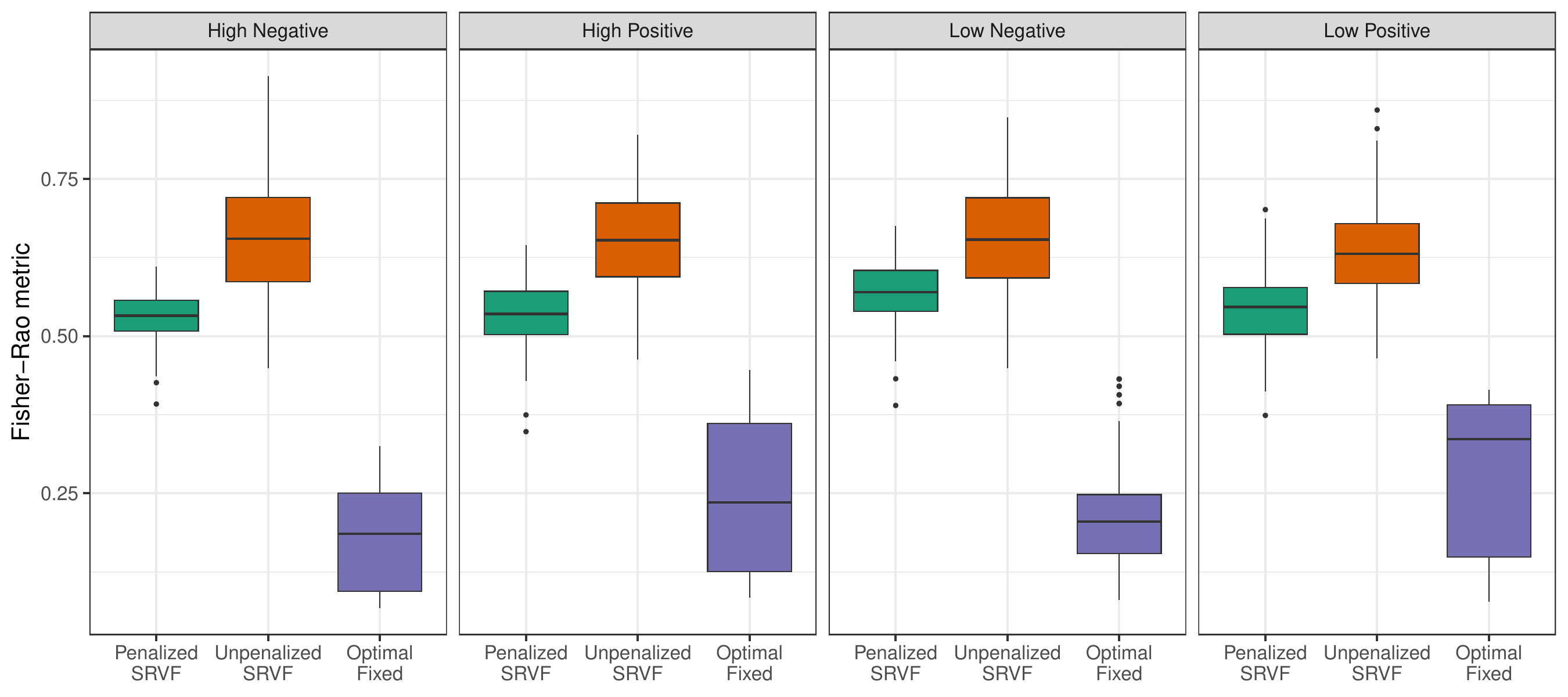}
  \caption{Boxplots for the estimated amount of warping, as measured by the Fisher-Rao metric between the identity warping $\gamma_{id}$ and the estimated warping function using unpenalized SRVF and penalized SRVF method with $\nu = 10s$ for each video. }
    
\end{figure}

\begin{figure}[H]
    \centering
        \includegraphics[width=\linewidth, page=4]{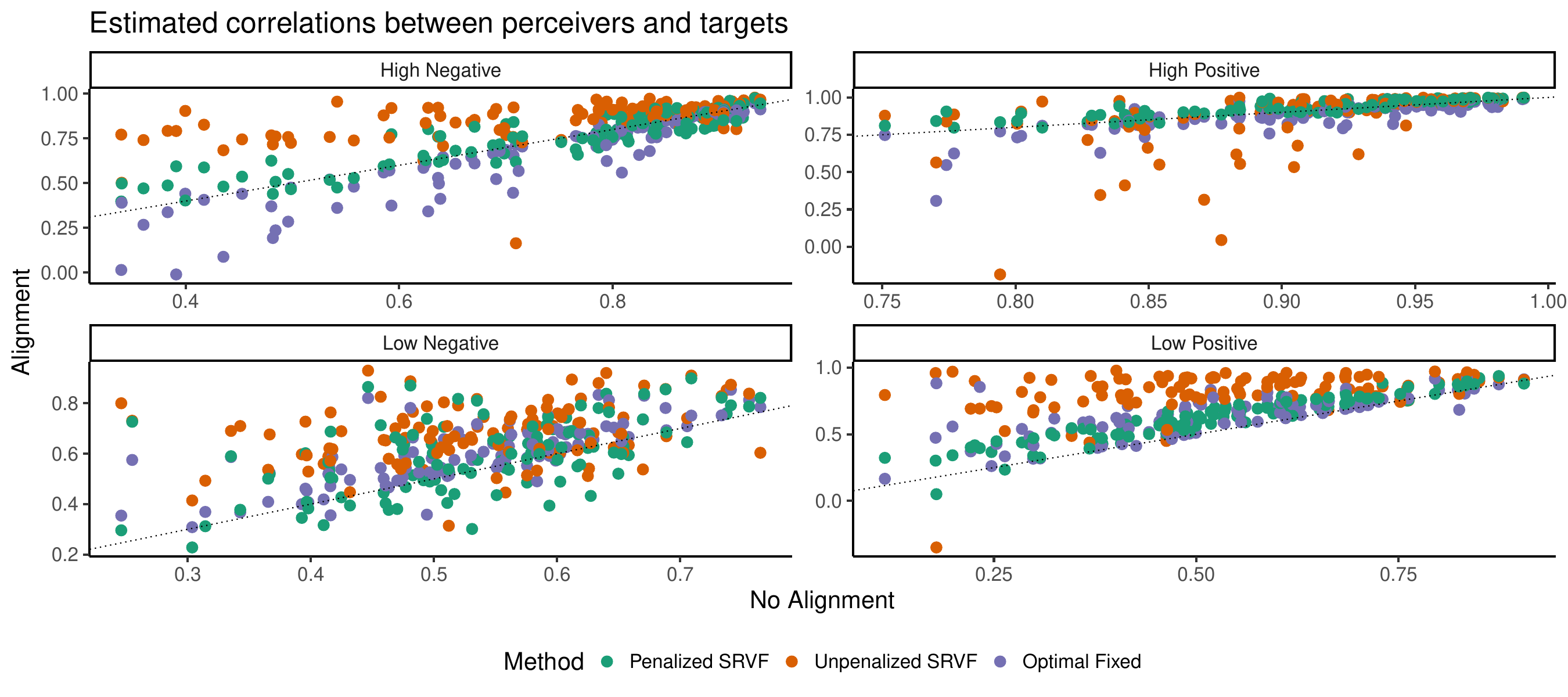}
    \caption{Scatterplots of correlation between target's ratings and each perceiver's ratings for each video. In each plot, the horizontal axis represents the correlation when no alignment is conducted, and the vertical axis represents the estimates under unpenalized SRVF (orange) and penalized SRVF with $\nu = 10s$ (green). }
  
\end{figure}
\FloatBarrier

\subsection{Data application: Music empathy}

This section contains additional results for the first data application, including the estimates of the amount of warping and all parameters under different thresholds.

\tabcolsep .3em
\begin{table}
\caption{Pairwise confidence intervals and $p$-values for the mean difference in the vertical distance and amount of warping obtained from different alignment methods in the music EA study, including no alignment, optimal fixed delay, unpenalized SRVF, and penaliszed SRVF with $\nu = 8s$.}
\begin{tabular}[t]{lp{1.5cm}lrccc}
\toprule
Music & Metric & Pair & Est & Low CI & Upp CI & $p$-value\\
\midrule
Anger & \multirow[t]{6}{1.2cm}{Amount of Warping} & Penalized SRVF - Unpenalized SRVF & -0.095 & -0.114 & -0.076 & 0.000\\
 &  & Penalized SRVF - Optimal Fixed & 0.312 & 0.289 & 0.334 & 0.000\\
 &  & Penalized SRVF - No Alignment & 0.545 & 0.534 & 0.557 & 0.000\\
 &  & Unpenalized SRVF - Optimal Fixed & 0.412 & 0.382 & 0.443 & 0.000\\
 &  & Unpenalized SRVF - No Alignment & 0.640 & 0.615 & 0.666 & 0.000\\
 &  & Optimal Fixed - No Alignment & 0.238 & 0.213 & 0.263 & 0.000\\
\addlinespace
 &  \multirow[t]{6}{1.2cm}{Vertical distance} & Penalized SRVF - Unpenalized SRVF & 0.094 & 0.061 & 0.127 & 0.000\\
 &  & Penalized SRVF - Optimal Fixed & -0.118 & -0.144 & -0.091 & 0.000\\
 &  & Penalized SRVF - No Alignment & -0.087 & -0.108 & -0.065 & 0.000\\
 &  & Unpenalized SRVF - Optimal Fixed & -0.212 & -0.251 & -0.173 & 0.000\\
 &  & Unpenalized SRVF - No Alignment & -0.182 & -0.218 & -0.147 & 0.000\\
 &  & Optimal Fixed - No Alignment & 0.031 & 0.016 & 0.047 & 0.000\\
\addlinespace
Happy & \multirow[t]{6}{1.2cm}{Amount of Warping} & Penalized SRVF - Unpenalized SRVF & -0.099 & -0.117 & -0.080 & 0.000\\
 &  & Penalized SRVF - Optimal Fixed & 0.209 & 0.186 & 0.232 & 0.000\\
 &  & Penalized SRVF - No Alignment & 0.526 & 0.518 & 0.535 & 0.000\\
 &  & Unpenalized SRVF - Optimal Fixed & 0.310 & 0.283 & 0.337 & 0.000\\
 &  & Unpenalized SRVF - No Alignment & 0.626 & 0.604 & 0.648 & 0.000\\
 &  & Optimal Fixed - No Alignment & 0.320 & 0.295 & 0.345 & 0.000\\
\addlinespace
 &  \multirow[t]{6}{1.2cm}{Vertical distance} & Penalized SRVF - Unpenalized SRVF & 0.277 & 0.192 & 0.362 & 0.000\\
 &  & Penalized SRVF - Optimal Fixed & -0.117 & -0.186 & -0.047 & 0.000\\
 &  & Penalized SRVF - No Alignment & -0.128 & -0.179 & -0.077 & 0.000\\
 &  & Unpenalized SRVF - Optimal Fixed & -0.389 & -0.467 & -0.311 & 0.000\\
 &  & Unpenalized SRVF - No Alignment & -0.410 & -0.487 & -0.332 & 0.000\\
 &  & Optimal Fixed - No Alignment & -0.019 & -0.067 & 0.029 & 0.290\\
\addlinespace
Sad & \multirow[t]{6}{1.2cm}{Amount of Warping} & Penalized SRVF - Unpenalized SRVF & -0.106 & -0.127 & -0.086 & 0.000\\
 &  & Penalized SRVF - Optimal Fixed & 0.271 & 0.251 & 0.290 & 0.000\\
 &  & Penalized SRVF - No Alignment & 0.518 & 0.509 & 0.528 & 0.000\\
 &  & Unpenalized SRVF - Optimal Fixed & 0.378 & 0.353 & 0.403 & 0.000\\
 &  & Unpenalized SRVF - No Alignment & 0.625 & 0.601 & 0.649 & 0.000\\
 &  & Optimal Fixed - No Alignment & 0.250 & 0.230 & 0.270 & 0.000\\
\addlinespace
 &  \multirow[t]{6}{1.2cm}{Vertical distance} & Penalized SRVF - Unpenalized SRVF & 0.052 & 0.039 & 0.066 & 0.000\\
 &  & Penalized SRVF - Optimal Fixed & 0.021 & 0.007 & 0.036 & 0.000\\
 &  & Penalized SRVF - No Alignment & -0.015 & -0.027 & -0.003 & 0.001\\
 &  & Unpenalized SRVF - Optimal Fixed & -0.032 & -0.052 & -0.012 & 0.000\\
 &  & Unpenalized SRVF - No Alignment & -0.069 & -0.088 & -0.050 & 0.000\\
 &  & Optimal Fixed - No Alignment & -0.037 & -0.046 & -0.029 & 0.000\\
\bottomrule
\end{tabular}
\end{table}

\begin{figure}
    \centering
    \includegraphics[width=\linewidth, page = 1]{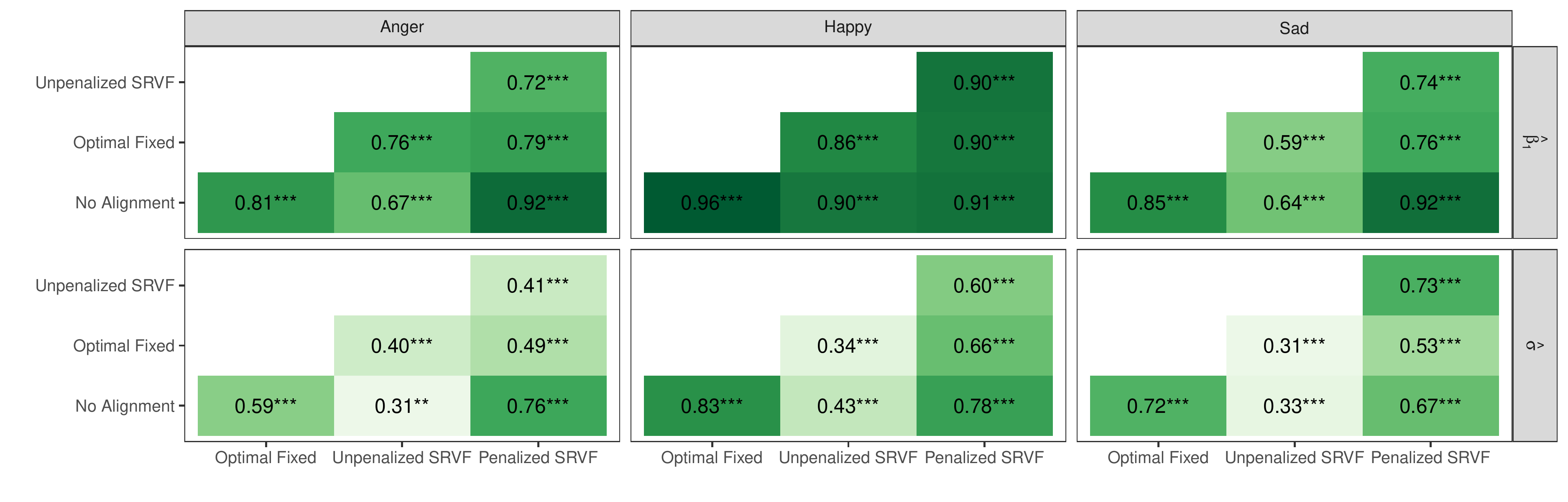}
    \caption{Correlation among EA measures (discrimination, $\hat\beta_1$ and standard deviation of random noise, $\hat\sigma$) in music EA study obtained by different alignment methods,  including no alignment, optimal fixed delay, unpenalized SRVF, and penaliszed SRVF with $\nu = 8s$}.
\end{figure}

\begin{figure}
    \centering
    \includegraphics[width=\linewidth, page = 1]{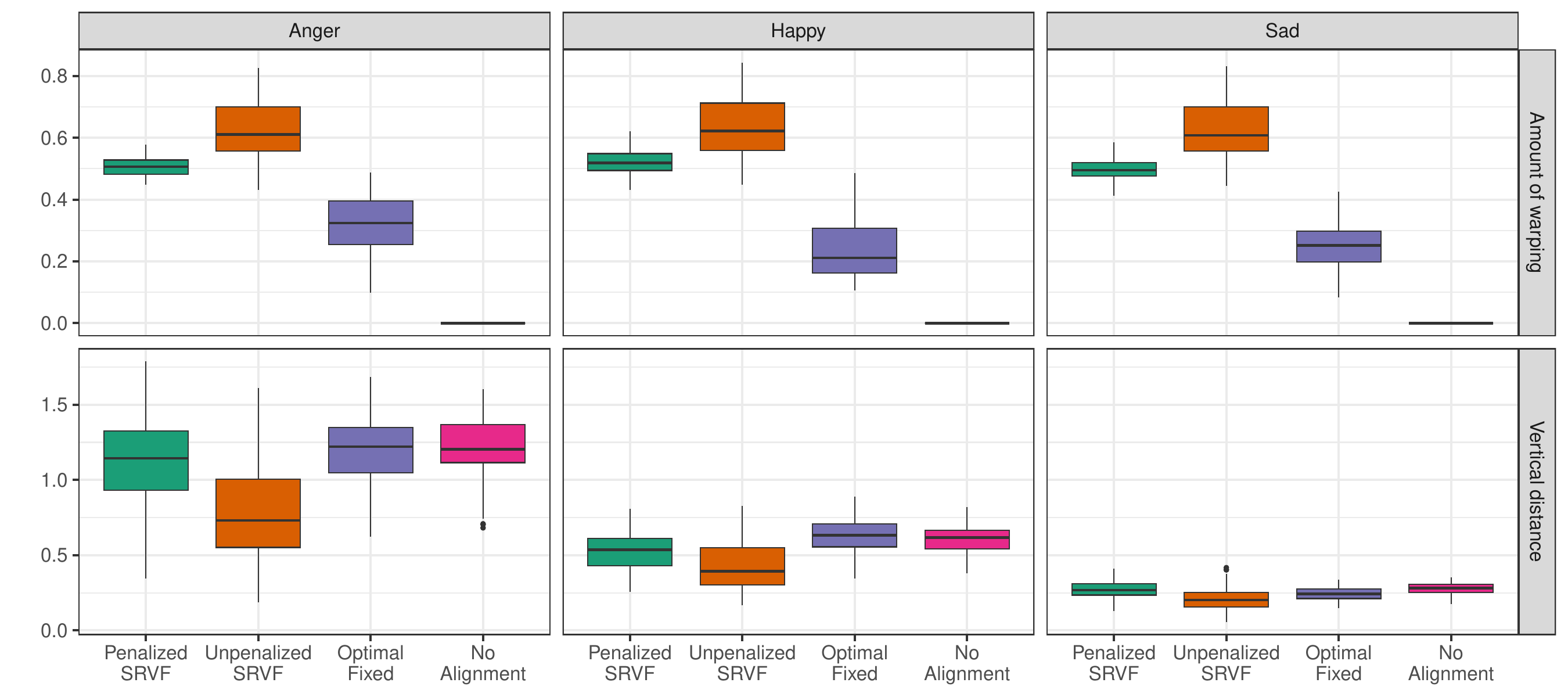}
    \caption{Boxplots of the vertical distance and the average amount of warpings of the estimated models for the three sets of music recordings. The penalized SRVF alignment was conducted using the threshold $\nu = 6s$.    }

\end{figure}

\begin{figure}[ht]
    \centering
    \includegraphics[width=\linewidth, page = 1]{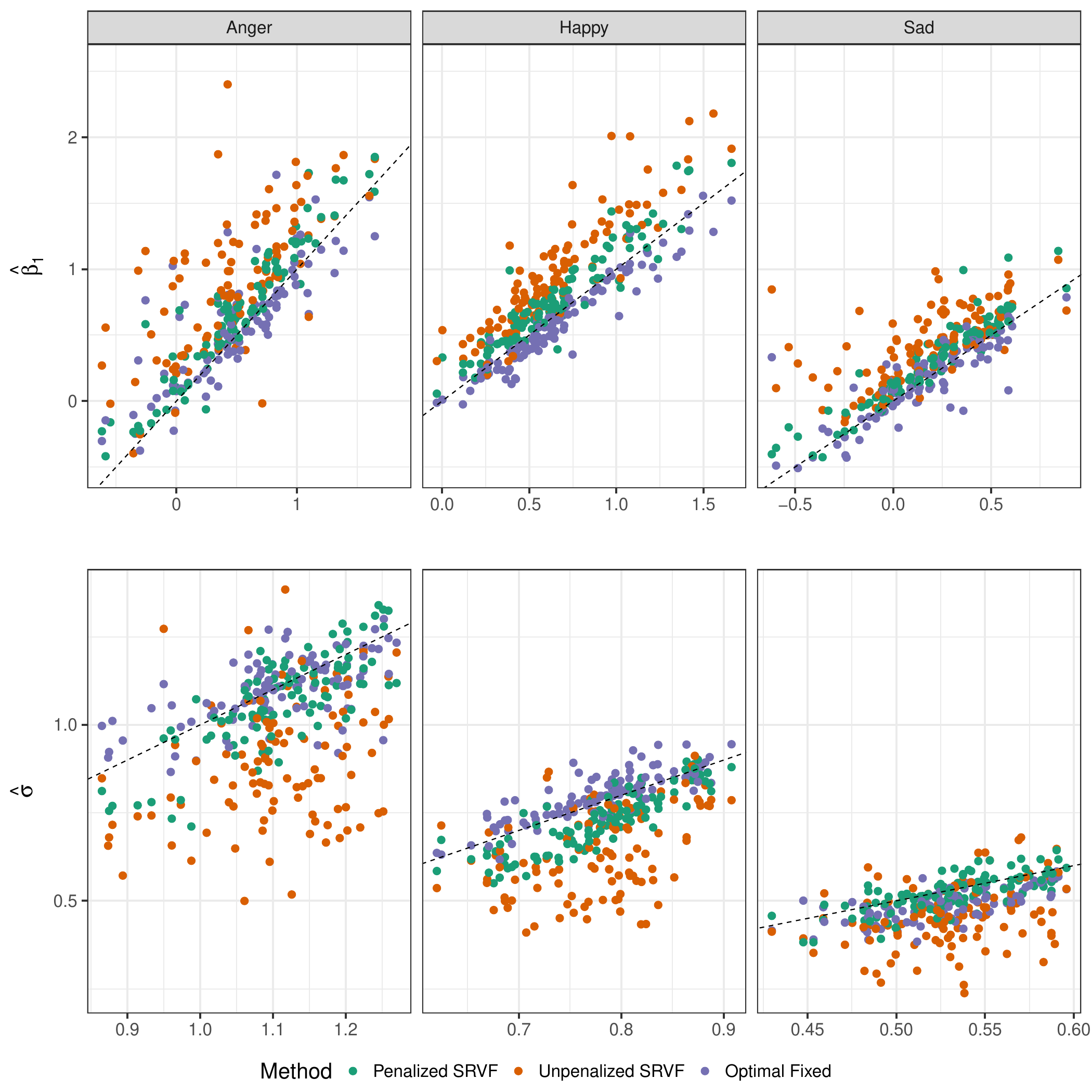}
    \caption{Scatterplots of estimates for the fixed effect discrimination $\beta$ (top) and random noise standard deviation $\sigma$ (bottom). In each plot, the horizontal axis represents the estimate when no alignment is conducted, and the vertical axis represents the estimates under unpenalized SRVF (orange) and penalized SRVF with $\nu = 6s$ (green). The dashed line represents the 45 degree line.   }
\end{figure}

\begin{figure}[ht]
    \centering
    \includegraphics[width=\linewidth, page = 1]{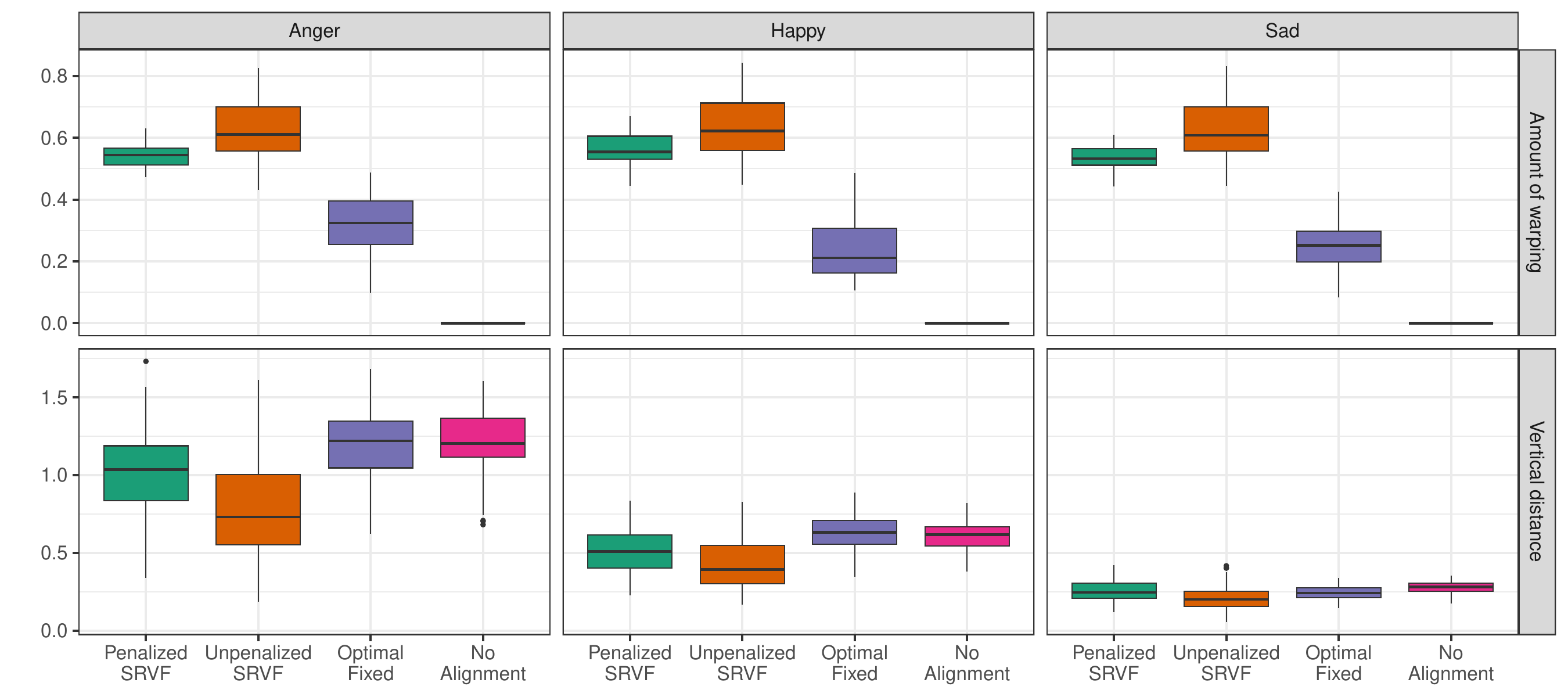}
    \caption{Boxplots of the vertical distance and the average amount of warpings of the estimated models for the three sets of music recordings. The penalized SRVF alignment was conducted using the threshold $\nu = 10s$.    }
\end{figure}
\FloatBarrier

\begin{figure}[ht]
    \centering
    \includegraphics[width=\linewidth, page = 1]{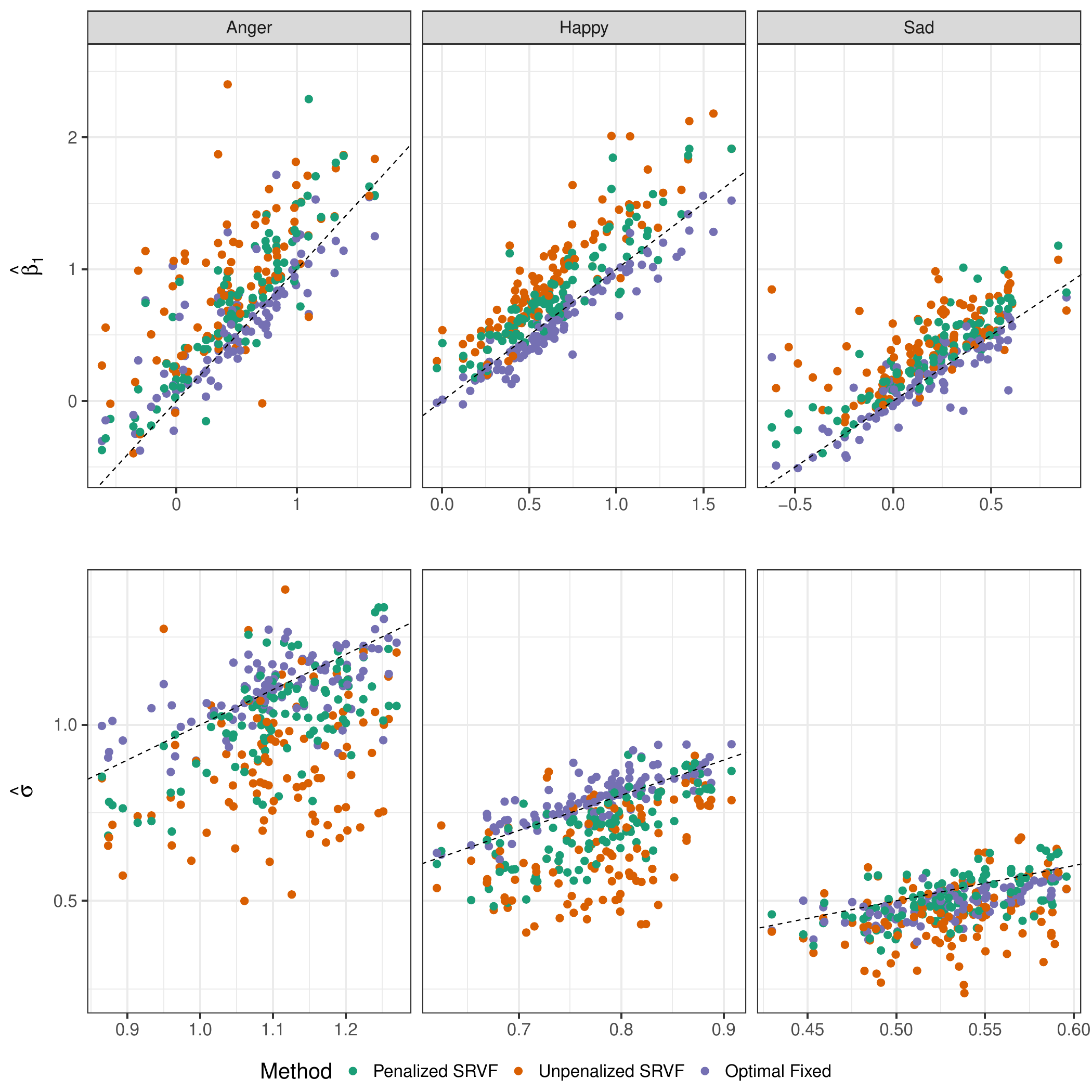}
    \caption{Scatterplots of estimates for the fixed effect discrimination $\beta$ (top) and random noise standard deviation $\sigma$ (bottom). In each plot, the horizontal axis represents the estimate when no alignment is conducted, and the vertical axis represents the estimates under unpenalized SRVF (orange) and penalized SRVF with $\nu = 10s$ (green). The dashed line represents the 45 degree line.   }

\end{figure}

\begin{table}
\centering
\caption{Mean (standard deviation) of parameter estimates across $113$ perceivers in the music empathy study using different alignment methods, including no alignment, unpenalized SRVF, and penalized SRVF with thresholds $\nu \in \{6, 8, 10\}$ seconds. }
\begin{tabular}{llllHlll}
\toprule
& Parameter &  \multirow{2}{1.5cm}{\centering No alignment} & \multirow{2}{1.5cm}{\centering Unpenalized  SRVF} & 
\multicolumn{4}{c}{Penalized SRVF} \\
& & & & \centering $\nu = 4$ & \centering $\nu = 6$& $\nu = 8$& $\nu = 10$ \\ 
\midrule
\multirow[t]{7}{*}{\raggedright\arraybackslash Anger} & Amount of warping & 0.00 (0.00) & 0.63 (0.09) & 0.48 (0.03) & 0.51 (0.03) & 0.53 (0.03) & 0.54 (0.04)\\

\addlinespace[0.5em]

& Vertical distance & 1.19 (0.24) & 0.80 (0.33) & 1.13 (0.31) & 1.10 (0.33) & 1.06 (0.32) & 1.01 (0.31)\\

\addlinespace[0.5em]
 & $\beta_0$ & 0.52 (3.75) & -1.32 (4.02) & -0.06 (4.12) & -0.26 (4.12) & -0.33 (4.11) & -0.51 (4.08)\\

\addlinespace[0.5em]
 & $\beta_1$ & 0.64 (0.81) & 1.08 (0.85) & 0.78 (0.89) & 0.83 (0.89) & 0.85 (0.89) & 0.89 (0.87)\\

\addlinespace[0.5em]
 
 & $\sigma$ & 1.09 (0.12) & 0.88 (0.19) & 1.05 (0.16) & 1.04 (0.17) & 1.02 (0.17) & 1.00 (0.16)\\

\addlinespace[0.5em]
 & $\sigma_a$ & 3.58 (3.10) & 3.63 (3.06) & 3.86 (3.22) & 3.94 (3.40) & 3.92 (3.24) & 3.98 (3.32)\\

\addlinespace[0.5em]

 & $\sigma_b$ & 0.81 (0.64) & 0.84 (0.62) & 0.87 (0.65) & 0.89 (0.69) & 0.89 (0.66) & 0.91 (0.67)\\

\addlinespace[1.5em]
\multirow[t]{7}{*}{\raggedright\arraybackslash Happy} & Amount of warping & 0.00 (0.00) & 0.64 (0.10) & 0.49 (0.04) & 0.52 (0.04) & 0.55 (0.05) & 0.56 (0.05)\\

\addlinespace[0.5em]
 & Vertical distance & 0.60 (0.10) & 0.42 (0.16) & 0.54 (0.12) & 0.53 (0.13) & 0.52 (0.14) & 0.51 (0.14)\\

\addlinespace[0.5em]
 & $\beta_0$ & 2.08 (2.95) & -0.27 (4.28) & 1.35 (3.03) & 1.22 (2.99) & 1.03 (2.99) & 0.92 (3.07)\\

\addlinespace[0.5em]
 & $\beta_1$ & 0.74 (0.53) & 1.13 (0.80) & 0.85 (0.56) & 0.87 (0.55) & 0.90 (0.55) & 0.91 (0.56)\\

\addlinespace[0.5em]

 & $\sigma$ & 0.77 (0.07) & 0.64 (0.12) & 0.73 (0.08) & 0.72 (0.09) & 0.71 (0.10) & 0.71 (0.10)\\

\addlinespace[0.5em]
 & $\sigma_a$ & 2.16 (1.89) & 3.02 (2.87) & 2.26 (1.92) & 2.17 (1.78) & 2.26 (1.79) & 2.43 (2.00)\\

 \addlinespace[0.5em]

 & $\sigma_b$ & 0.34 (0.35) & 0.50 (0.59) & 0.36 (0.37) & 0.35 (0.33) & 0.37 (0.32) & 0.39 (0.37)\\[1em]

\addlinespace[0.5em]
\multirow[t]{7}{*}{\raggedright\arraybackslash Sad}& Amount of warping & 0.00 (0.00) & 0.63 (0.09) & 0.47 (0.03) & 0.50 (0.04) & 0.52 (0.04) & 0.53 (0.04)\\

\addlinespace[0.5em]
 & Vertical distance & 0.28 (0.04) & 0.21 (0.08) & 0.27 (0.05) & 0.27 (0.06) & 0.26 (0.07) & 0.26 (0.07)\\

\addlinespace[0.5em]
 & $\beta_0$ & 3.10 (1.91) & 2.13 (3.40) & 2.73 (1.93) & 2.61 (2.00) & 2.47 (2.04) & 2.35 (2.08)\\

\addlinespace[0.5em]
 & $\beta_1$ & 0.21 (0.41) & 0.43 (0.69) & 0.30 (0.41) & 0.32 (0.42) & 0.36 (0.43) & 0.38 (0.44)\\

\addlinespace[0.5em]
 
 & $\sigma$ & 0.53 (0.04) & 0.45 (0.09) & 0.52 (0.05) & 0.52 (0.06) & 0.51 (0.06) & 0.50 (0.07)\\

\addlinespace[0.5em]
 & $\sigma_a$ & 2.49 (2.51) & 2.85 (5.17) & 2.54 (2.87) & 2.60 (3.01) & 2.62 (3.00) & 2.70 (3.21)\\
\addlinespace[0.5em]
 & $\sigma_b$ & 0.49 (0.50) & 0.56 (1.04) & 0.50 (0.58) & 0.51 (0.61) & 0.51 (0.61) & 0.53 (0.65)\\
\bottomrule
\end{tabular}
\end{table}

\end{document}